\documentclass{aastex62}

\received{September ??, 2018}
\revised{, 2018}
\accepted{December ??, 2018}%\today}
\submitjournal{ApJ}

\shorttitle{Inner star formation in local MIR bright QSOs}
\shortauthors{Mart\'inez-Paredes et al.}
\begin{document}

\title{Quantifying star formation activity in the inner 1kpc of local MIR bright QSOs}

\correspondingauthor{M. Mart\'inez-Paredes}
\email{mariellauriga@kasi.re.kr, m.martinez@irya.unam.mx}

\author[0000-0002-0786-7307]{M. Mart\'inez-Paredes}
\affil{Instituto de Radioastronom\'ia y Astrof\'isica UNAM 
Apartado Postal 3-72 (Xangari), 58089 Morelia, Michoac\'an, Mexico}  
\affil{Korea Astronomy and Space Science Institute 776, Daedeokdae-ro, Yuseong-gu, Daejeon, Republic of Korea (34055)}

\author{I. Aretxaga}
\affiliation{Instituto Nacional de Astrof\'isica, \'Optica y Electr\'onica (INAOE), Luis Enrrique Erro 1, Sta. Ma. Tonantzintla, Puebla, Mexico}
%\collaboration{(AAS Journals Data Scientists collaboration)}

\author{O. Gonz\'alez-Mart\'in}
\affil{Instituto de Radioastronom\'ia y Astrof\'isica UNAM 
Apartado Postal 3-72 (Xangari), 58089 Morelia, Michoac\'an, Mexico}  

\author{A. Alonso-Herrero}
\affiliation{Centro de Astrobiolog\'ia, CSIC-INTA, ESAC Campus, E-28692 Villanueva de la Cañada, Madrid, Spain}

\author{N. A. Levenson}
\affiliation{Space Telescope Science Institute, Baltimore, MD 21218, USA}

\author{C. Ramos Almeida}
\affiliation{Instituto de Astrof\'isica de Canarias (IAC),e-38205 La Laguna, Tenerife, Spain}
\affiliation{Departamento de Astrof\'isica, Universidad de La Laguna (ULL), E-38206 La Lagun, Tenerife, Spain}
\author{E. L\'opez-Rodr\'iguez}
\affiliation{Stratospheric Observatory For Infrared Astronomy (SOFIA), NASA Ames Research Center, CA, USA 94041}



\begin{abstract}

 We examine star formation activity in a distance- ($z<0.1$) and flux-limited sample of quasars (QSOs). Mid-infrared (MIR) spectral diagnostics at high spatial resolution ($\sim0.4$ arcsec) yield star formation rates (SFRs) in the inner regions ($\sim$300 pc to 1 kpc) for 13 of 20 of the sample members. We group these objects
according to the size probed by the high angular resolution spectroscopy, with characteristic scales of $<0.7$ and $\sim0.7-1$ kpc. Using the polycyclic aromatic hydrocarbon (PAH) feature at 11.3 $\mu$m, we measure SFRs around 0.2 and 1.6 M$_{\odot}$yr$^{-1}$. We also measure the larger aperture PAH-derived SFRs in the individual {\it IRS/Spitzer} spectra of the sample and obtain a clear detection in $\sim58$ percent of them. We compare smaller and larger aperture measurements and find that they are similar, suggesting that star formation activity in these QSOs is more centrally concentrated, with the inner region ($\lesssim1$ kpc) accounting for the majority of star formation measured on these scales, and that PAH molecules can be present in most local MIR-bright QSOs within a few hundred pc from the central engine.
 By comparison with merger simulations, we find that our estimation of the SFR and black hole 
(BH) accretion rates are consistent with a scenario in which the star formation activity is 
centrally peaked as predicted by simulations.

\end{abstract}



\section{Introduction} \label{sec:intro}

An active galactic nucleus (AGN) harbors a supermassive black hole (SMBH, $M_{BH}\sim10^{6}-10^{9}$ M$_{\odot}$), which is surrounded by an accretion disk responsible for the strong radiation field that ionizes the gas around it \citep[e.g.,][]{Magorrian98, Peterson00, Gultekin09, Du15}. Several studies have postulated that some of the gas in the vicinity of the AGN feeds the SMBH and fuels a central starburst located in comparable scales to a dusty torus \citep[e.g.,][]{CidFernandez_Terlevich95, Gonzalez98, Wada02, Thompson05, Ballantyne08, Diamond10, Feruglio10, Miller15}. Indeed, there is observational evidence of nuclear (few hundred pc scales) starbursts in several narrow-line AGN, such as Seyfert 2 and low-ionization nuclear emission-line regions (LINERs), which have been detected through a variety of methods and wavelengths \citep[e.g.,][]{E.Terlevich92, Oliva95, Colina97, Heckman97, Gonzalez98, Povic16}, and in Seyfert 1 galaxies \citep[e.g.,][]{Imanishi_Wada04, Deo06}. Despite the large effort to understand how AGN activity influences the central star formation, this is still an open issue. Some studies have found a positive correlation between the star formation rate (SFR) and AGN luminosity \citep[e.g.,][]{Netzer09b, Delvecchio15, Matsuoka15, Gurkan15, Dong_Wu16, Esquej14} while others have found none \citep[e.g.,][]{Rosario12, Azadi15, Stanley15, Shimizu15}. One explanation for this might be the different physical scales and AGN luminosities traced by the various works, as well as finding good estimators of the SFR in AGN.

Finding evidence of starbursts around type-1 AGN has proven difficult, since the bright nucleus outshines classical starburst features, like the UV continuum emission and optical or near-infrared emission lines \citep[e.g.,][]{E.Terlevich90, Colina97, Voit92, Cresci04, Davies07}. Using IRS/{\it Spitzer} observations of  Palomar-Green (PG) QSOs with a redshift $z<0.5$, \citet{Shi07} detected polycyclic aromatic hydrocarbons (PAHs) at 7.7 and 11.3 $\mu$m against the strong MIR AGN continuum on scales of $\sim2-20$ kpc.
On the other hand, measuring the equivalent width (EW) of the PAH at 11.3 $\mu$m in the  QUEST {\it Spitzer} survey, \citet{Schweitzer08} found evidence for star formation in a sample of 27 PG QSOs at $z<0.3$; 40 percent of objects showed clear PAH features, and for those that lacked individual detections the stacked spectrum revealed them, implying that starbursts are present in most QSOs at $\simeq3.6$ arcsec spatial resolution ($\sim3-15$ kpc).

There is a large variety of methods to estimate the SFR from PAHs \citep[][]{Treyer10, Diamond12, Pope08, Farrah07}, and the results could vary up to a factor of two according to the size of the sample and method used \citep{Shipley16}. \citet{Shi14} found a tight correlation between the SFR derived from the PAHs at 11.3 $\mu$m and the far-IR luminosity, suggesting that the PAH emission at 11.3 $\mu$m is a good indicator of the star formation activity in QSOs \citep[see also,][]{Shi07, Netzer07}. Additionally, \citet{Shipley16} calibrated the luminosity of the PAH features at 6.2, 7.7, and 11.3 $\mu$m as a measure of the SFR, and showed that the PAH SFR method is as accurate as those based on hydrogen recombination lines (i.e., $H_{\alpha}$, Pa$\alpha$). 

High angular resolution studies of nuclear ($<100$ pc) star formation in AGN in the MIR  are still scarce and limited to nearby Seyfert galaxies \citep[e.g.,][]{Roche06, Mason07, Watabe08, Gonzalez13, A-AH14, Esquej14, Ramos14, Ruschel-Dutra17, Esparza17}. 
These studies showed that PAH emission in the nuclear region (few tens of pc) can be explained by the ionization produced by nuclear starbursts. Nevertheless, a recent study suggests that AGN excitation could also be an important component in heating/exciting PAH molecules in the nuclear region of active galaxies \citep{Jensen17}.

Here we use MIR high angular resolution spectra to estimate, for the first time, the star formation activity within the central several hundred pc in local bright MIR QSOs using the PAH at 11.3 $\mu$m. We compare our results with merger simulations presented by \citet[][]{Hopkins_and_ Quataert10}, which predict a relationship between the SFR, on few pc and several tens of kpc, and the activity of the AGN. The paper is organized as follows: Section \ref{sample} describes the sample and observations, in 
section \ref{analysis} we present our measurements of the PAH feature, in sections \ref{SFR} and \ref{discussion} we discuss our results and in section \ref{conclusion} we present the conclusions. Throughout this paper 
we have assumed the following cosmology: $H_{0}=70$ km s$^{-1}$Mpc$^{-1}$, $\Omega_{m}=0.3$, $\Omega_{\Lambda}=0.7$.

\begin{table*}
	\begin{minipage}{1.\textwidth}
		\caption{Main properties and observational details of the QSO sample. Columns 1 and 2 give the name and assigned number group (defined in section \ref{nuc_pah}), 
columns 3, 4 and 5 the redshift, angular scale, and distance, column 6 lists the intrinsic hard X-ray (2-10 keV) luminosity, and columns 7 and 8 the inner physical scale, probed by the slit-width of the high angular resolution spectrum, and the instrument and telescope used to obtain the high angular resolution spectrum. \label{tab:basic1}}
\centering
\begin{tabular}{lccccccc}
				\hline
				Name &Group 	&$z^{a}$  & Angular scale   & d &$L^{b}_{X (2-10 {\rm keV})}$ &Inner Physical Scale &Nuclear Spectrum$^{c}$   \\
				           &          &      & (kpc/arcsec)&    (Mpc) &(erg s$^{-1}$)  & pc &  \\  
				\hline
				PG~1501+106/MRK~841   & 1& 0.0364     &  0.723  &  160 & 7.$8\times10^{43}$    & 376 & CC/GTC \\
				MRK~509	                          &1 &0.0344     &  0.685 &  151 & 4.$8\times10^{44}$    &  514&  VISIR/VLT  \\
				PG~2130+099/IIZw136      &1 & 0.0630     &  1.213 &  283 &  3.$2\times10^{43}$     & 631 & CC/GTC \\
				PG~1229+204/MRK~771    &1 & 0.0630     &  1.213  &  283 &3.$1\times10^{43}$    &  631& CC/GTC \\
				PG 0844+349	                   & 1& 0.0640     &  1.231  &  287 &  5.$5\times10^{43}$    & 640 &  CC/GTC \\
				MR~2251-178                   & 1& 0.0640     &     1.231        &  287 & 2.$9\times10^{44}$     &    640       & CC/GTC\\
			    PG~0003+199/MRK~335  & 1 & 0.0258     &  0.519 &  113  & 1.$9\times10^{43}$    &  670& CC/GTC \\
			    PG~1440+356/MRK~ 478  &  2& 0.0791      & 1.494  &  359 & 5.$8\times10^{43}$   & 777 & CC/GTC  \\
			    PG 1211+143                      &2 & 0.0809     & 1.525   &  368 & 5.$0\times10^{43}$    & 793 &  CC/GTC \\
			    PG~1426+015/MRK~1383  & 2& 0.0866     & 1.622  &  395 & 1.$3\times10^{44}$    &  843& CC/GTC \\
			    PG~1411+442 	                   &2 & 0.0896     & 1.627  &  397 & 2.$5\times10^{43}$    &  846 & CC/GTC  \\
			    PG~0050+124/IZw1           &2 & 0.0589     & 1.139 &  264 &  7.$1\times10^{43}$    & 854 & VISIR/VLT \\
			    PG 0804+761	                   & 2& 0.1000     &1.844   &  460 &2.$9\times10^{44}$    & 959  & CC/GTC \\
				PG~1448+273                    & ... & 0.0650      &  1.248     &  292    &  2.$0\times10^{43}$ &   ...         & ...\\ 			    PG~1534+580                    & ...& 0.0296      &   0.593       &  130    & 1.$8\times10^{43}$ &  ...        &  ...\\
				PG~1535+547                    &... & 0.0389     &   0.771       & 172      &  4.$0\times10^{42}$ &     ...      & ...\\				PG~2214+139                    &... & 0.0658     &   1.263       &  296     &  6.$6\times10^{43}$ &     ...      & ....\\
				PG~0923+129                     &... & 0.0292     & 0.585 & 128      &  2.$6\times10^{43}$ &  ...        &   ...\\
				PG~1351+640                     &... & 0.0882     &   1.649      &  403      & 1.$2\times10^{43}$ &    ...        & ...  \\			    PG~0007+106/MRK~1501 &... & 0.0893    & 1.667  &  408    & 1.$4\times10^{44}$ &  ...         & ... \\

				\hline
			\end{tabular}\\
			{\bf References}.$^{a}$NED, $^{b}$\citet{Zhou2010},  $^{c}$\citet{M-MP17} and references therein.\\			
		\end{minipage}
	\end{table*}	

\section{Sample and Observations}
\label{sample}
We use the sample of \citet{M-MP17} which was 
built according to the following criteria: 1) a redshift $z<0.1$ in order to study the inner MIR emission at scales $\lesssim1$ kpc; 2) an N-band flux 
$f_{N}\geq0.02$ Jy in order to detect these objects from the ground; and 3) a hard X-ray luminosity $L_{2-10\, \text{keV}}>10^{43}$ erg s$^{-1}$ as an 
 indicator of intrinsic powerful AGN activity. This sample of 20 AGN is representative of nearby, MIR-bright quasars.

These data enable us to study the inner (few hundred pc) MIR emission of the PAH at 11.3 $\mu$m in QSOs, that is directly associated with recent star formation. Table~\ref{tab:basic1} lists basic galaxy properties of the sample. 

\begin{figure*}
%∫∫\hspace{-1cm}
\begin{tabular}{cc}
%\centering
	\includegraphics[scale=0.39]{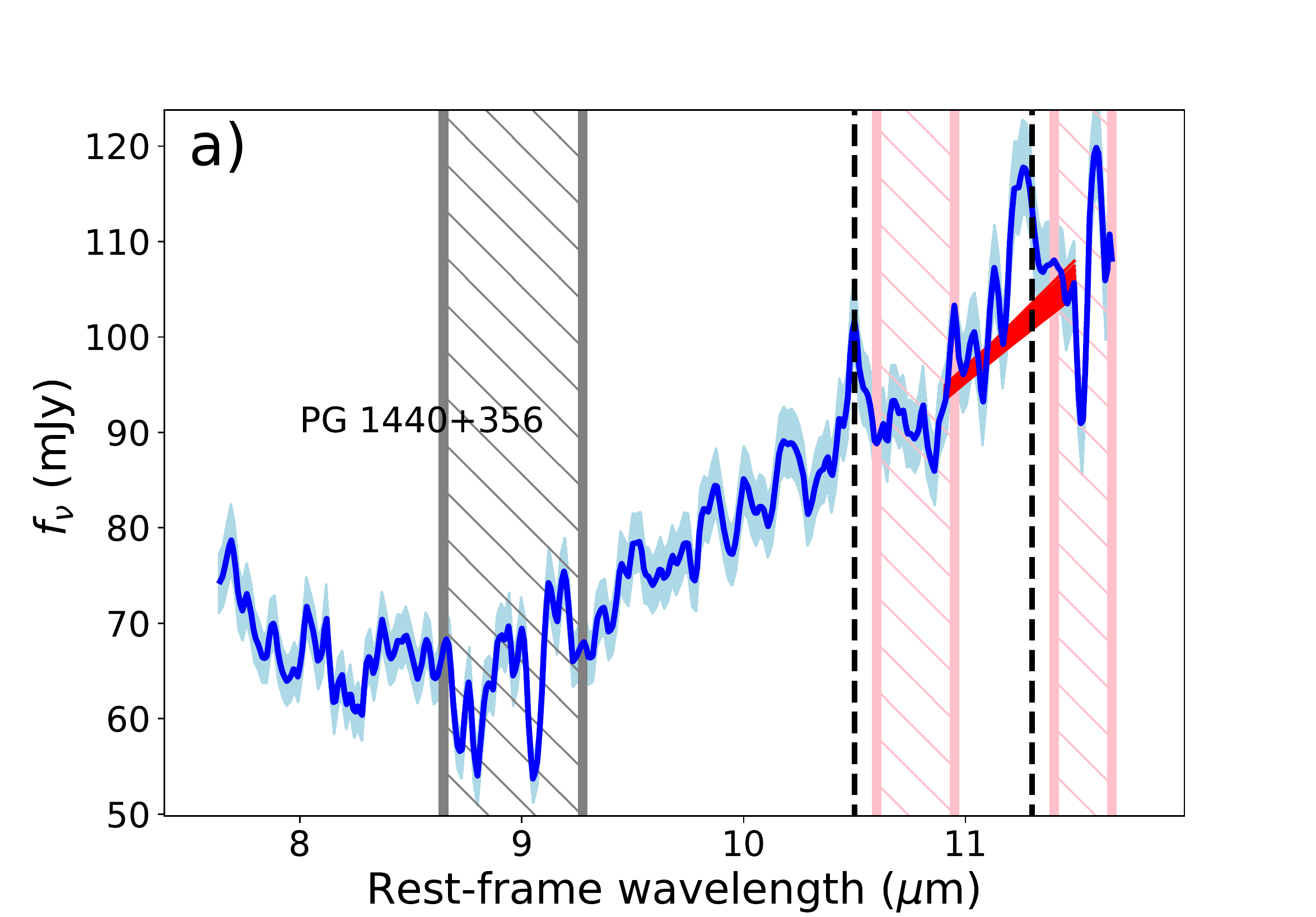} &\includegraphics[scale=0.35]{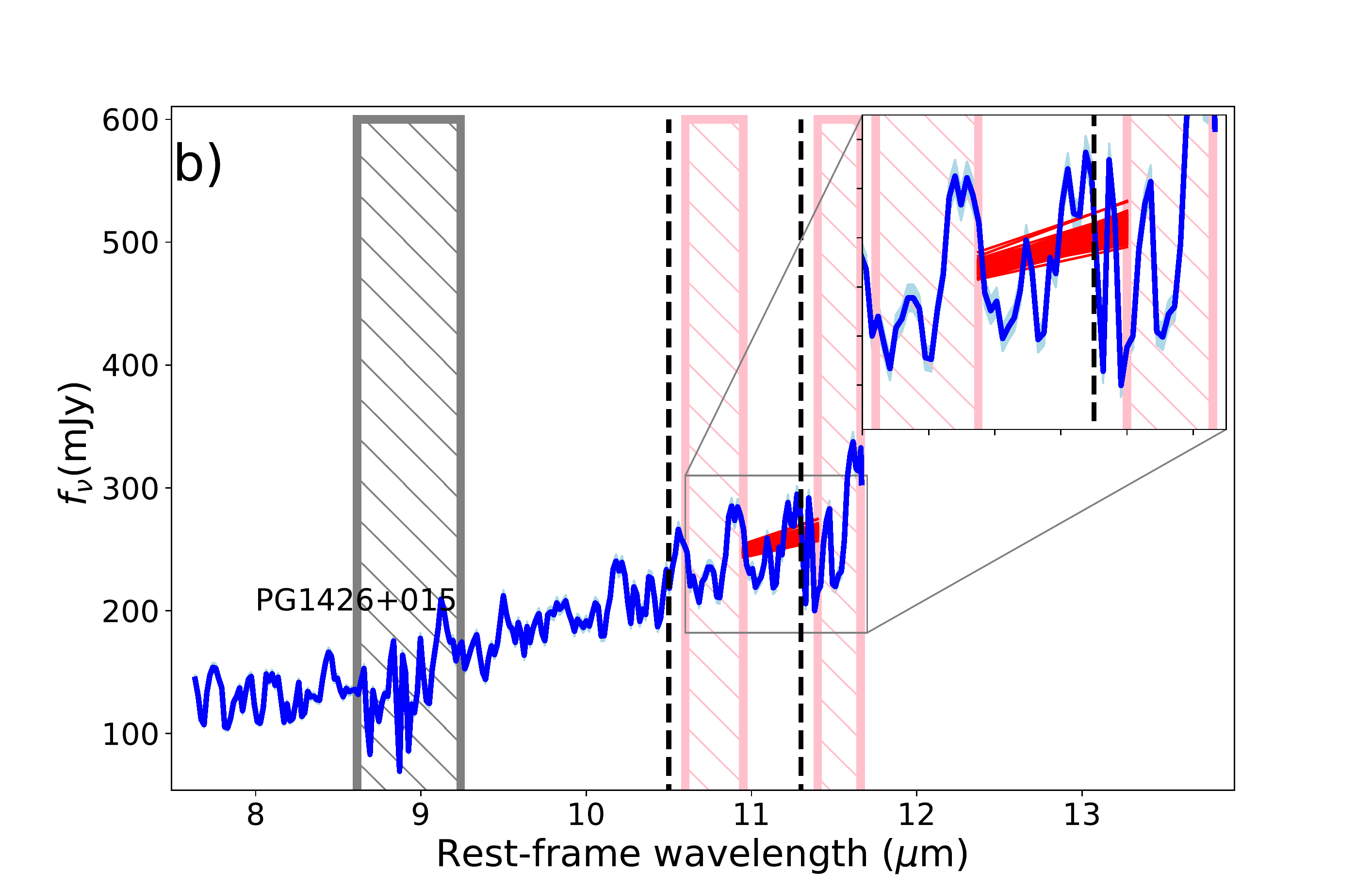}\\
	\multicolumn{2}{c}{\includegraphics[scale=0.33]{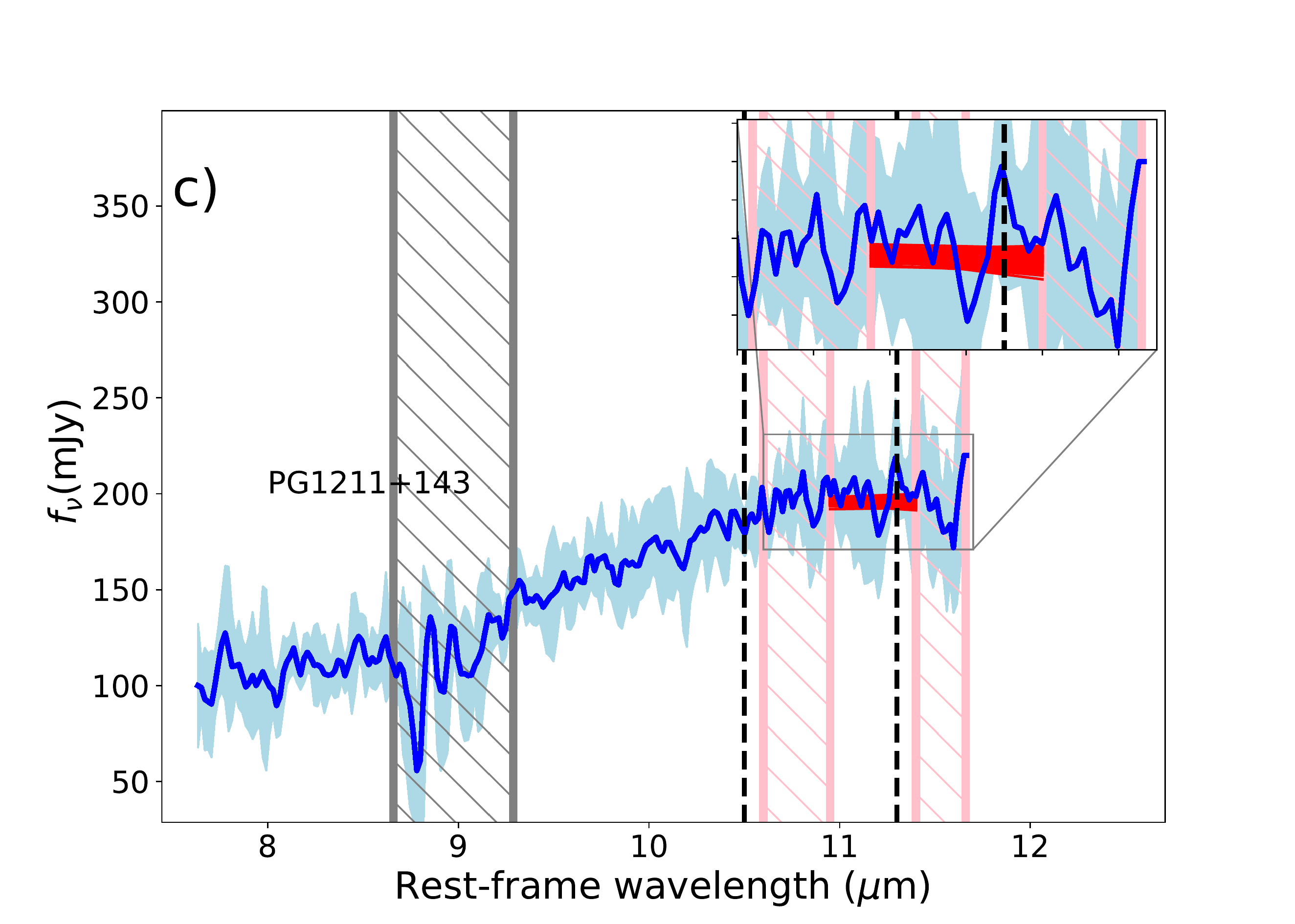}}\\
\end{tabular}	
    \caption{{\bf CC spectra} of PG~1440+356 (MRK~478, panel a), PG~1426+015 (panel b), and PG~1211+143 (panel c). The solid red lines are local continua generated through Monte Carlo simulations. The vertical dashed lines indicate the position of the PAH at 11.3 $\mu$m and of the [S IV]10.5$\mu$m emission. The two hatched pink regions represent the bands used to measure the continua. The grey hatched region marks the approximate spectral range of low atmospheric transmission.}
    \label{ind_nuc}
\end{figure*}

\subsection{Inner ($\lesssim1$ kpc) N band spectra}
We have inner ($\lesssim1$ kpc) spectra in the N band ($\sim7.5-13.5\mu$m, rest-frame wavelength) of 13 QSOs. Eleven of these were obtained with the MIR instrument CC on GTC as part of the ESO-GTC program \citep[P.I: A. Alonso-Herrero, ID program: 182.B-2005,][]{A-AH16a}, guaranteed time \citep[P.I: C. Packham,][]{Packham05a} program, and Mexican open time (P.I.: I. Aretxaga/M. Mart\'inez-Paredes). The spectra were obtained with a slit-width of 0.52 arcsec in low resolution mode ($R=175$). The data were reduced using the pipeline for CanariCam developed by \citet{Gonzalez13}. For more details on the observations and data reduction please refer to \citet{M-MP17} and \citet[][]{A-AH16a, A-AH16b}. 

Two of the MIR spectra were obtained with VISIR on the VLT, MRK~509 \citep[][]{Hoenig10} and PG~0050+124 \citep[][]{Burtscher13}. These spectra were obtained in low spectral resolution mode ($R\sim300$), with a slit-width of 0.75 arcsec and angular resolution $\sim0.3$ arcsec.

\subsection{\emph{IRS/Spitzer} spectra ($\lesssim6$ kpc)}
We obtained the reduced 2D low resolution ($R\sim60-127$) \emph{IRS/Spitzer} spectra \citep[][]{Schweitzer06, Shi07} for all 19 QSOs available in the CASSIS database \citep[v6.,][]{Lebouteiller11}. MR~2251-178 does not have a {\it Spitzer} spectrum. The spectra include the SL1 ($\lambda\sim7.4-14.5\,\mu$m) and SL2 ($\lambda\sim5.2-7.7\,\mu$m) modules acquired with a slit-width of 3.6 arcsec, and the LL1 ($\lambda\sim19.9-39.9\mu$m) and LL2 ($\lambda\sim13.9-21.3\,\mu$m) modules acquired with a slit-width of 10.5 arcsec \citep[][]{Werner04, Houck04}. We built stitched spectra between $5-35$ $\mu$m by scaling the flux of the LL and SL1 modules in the overlaping wavelengths of module SL2. The scaling factors were around 1.1 \citep{M-MP17}. 

\begin{figure*}
\begin{tabular}{cc}
\hspace{-1cm}
\includegraphics[scale=0.36]{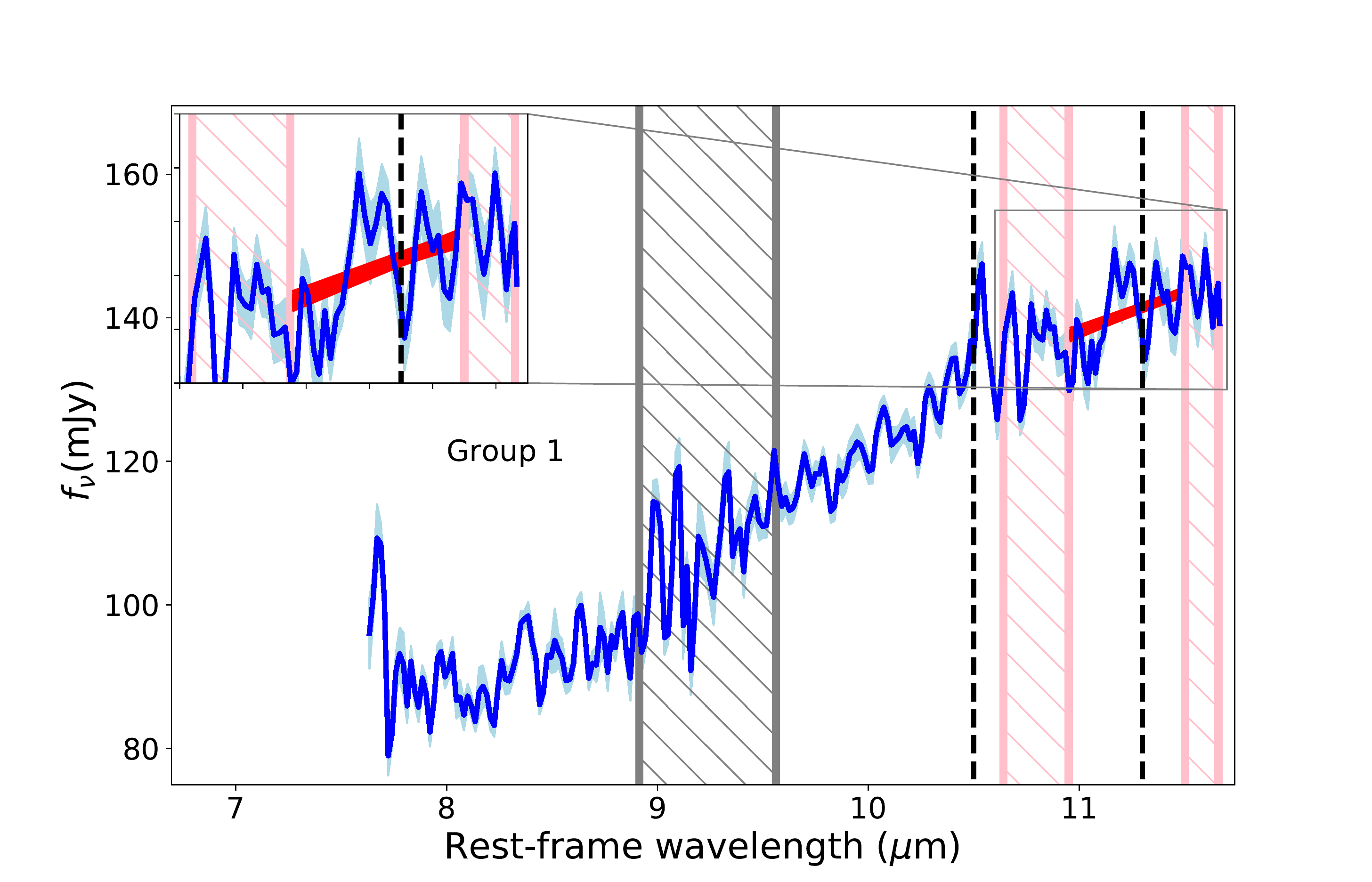} &\hspace{-1cm}\includegraphics[scale=0.36]{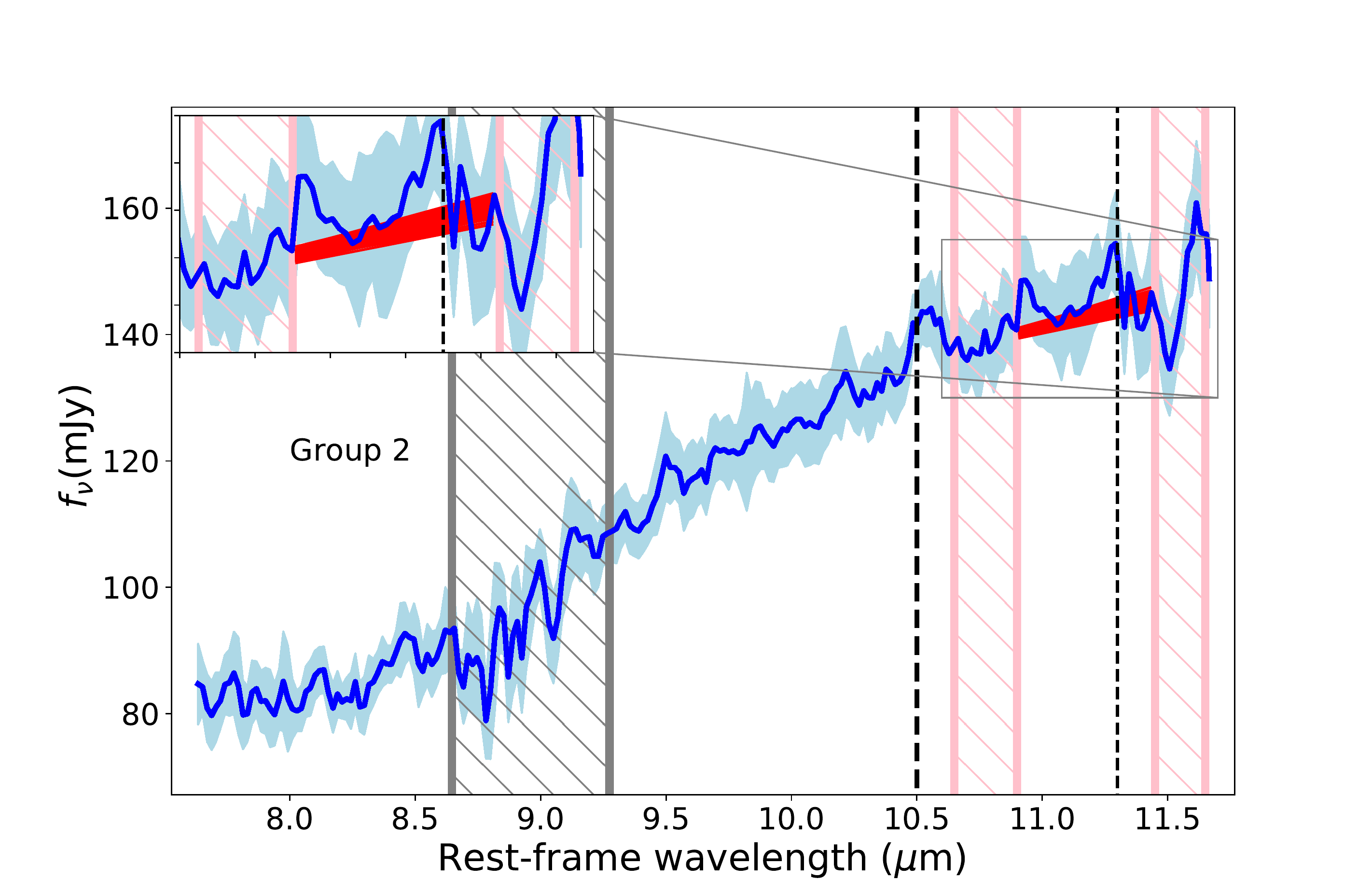}\\ 
 \end{tabular}
  \caption{Stacked rest-frame high angular resolution spectra. Left: stacked spectrum of group 1 (physical scale $<0.7$ kpc). Right: stacked spectrum of group 2 (physical scale $<1$ kpc). The solid red lines are the local continua generated through Monte Carlo simulations. The vertical dashed black lines indicate the position of the PAH at 11.3 $\mu$m and of the [S IV]10.5$\mu$m emission. The two hatched pink regions show the bands used to measure the continua. The grey hatched region marks the approximate spectral range of low atmospheric transmission.}
 \label{stack_nuc}
\end{figure*}

\begin{table}
\caption{{\bf Inner (CC and VISIR) and larger ({\it Spitzer})} aperture flux and EW of the PAH at 11.3 $\mu$m, plus the SFR and BH accretion rate. Column 1 gives the name, columns 2, 3 and 4 give the inner aperture flux, EW, and SFR, while columns 5, 6 and 7 give the larger aperture flux, EW, and SFR. Column 8 lists the BH accretion rates. \label{tab:_indv_pah}}	
	\begin{minipage}{1.\textwidth}
\centering
%\scalebox{.9}{
\begin{tabular}{lccccccc}
\hline
    & \multicolumn{3}{c}{Inner aperture} & \multicolumn{3}{c}{Larger aperture} & \\
				\hline
				Name   & $f_{11.3\,\mu\text{m}}$ & EW$_{11.3\,\mu\text{m}}$  & SFR$_{IRS}$ & $f_{11.3\,\mu\text{m}}$ & EW$_{11.3\,\mu\text{m}}$ & SFR$_{inner}$ &$\dot{M}_{BH}$\\
			           & $10^{-13}$  erg s$^{-1}$cm$^{-2}$& $\mu$m&$M_{\odot}yr^{-1}$  &  $10^{-13}$  erg s$^{-1}$cm$^{-2}$     &$\mu$m  & $M_{\odot}yr^{-1}$  &$M_{\odot}yr^{-1}$ \\  
				\hline
				\hline	
	
				 PG~1501+106 & $<0.23$  & $<0.057$ &  $<0.2$ & $<0.3$ &  $<0.007$  &    $<0.2$ & 0.1     \\			     MRK~509     & $1.15\pm0.07$ & $0.014\pm0.001$ &   $0.7\pm0.2^{*}$ & $2.65\pm0.01$  &  $0.0401\pm0.0001$  &   $1.7\pm0.5$   &1.1      \\
				 PG~2130+099 & $0.38\pm0.04$  & $0.028\pm0.002$ & $0.8\pm0.3$ & $0.45\pm0.02$ &  $0.0105\pm0.0004$  &    $1.0\pm0.3$   &0.04    \\
				 PG~1229+204 & $<0.3$ & $<0.02$ &  $<0.6$ & $<0.2$ &  $<0.02$  &   $<0.5$   & 0.04     \\
				 PG~0844+349 &  $<0.2$ & $<0.01$ &  $<0.4$ & $<0.2$ &  $<0.02$  &   $<0.5$     & 0.1     \\
				 MR~2251-178 &  $<0.54$ & $<0.03$ &       $<1.6$      &   ...  & ...   & ...               &  0.6    \\
				 PG~0003+199 & $0.48\pm0.05$ & $0.022\pm0.003$ &  $0.15\pm0.05$  & $0.493\pm0.005$ & $0.0109\pm0.0001$   & $0.15\pm0.05$  & 0.002 \\
				 PG~1440+356 &  $0.83\pm0.07$ & $0.035\pm0.003$&   $3\pm1$ & $1.607\pm0.008$ &  $0.0739\pm0.0004$  &    $6\pm2$  &0.1     \\
				 PG~1211+143 &  $<0.4$  & $<0.02$ &     $<1.5$  & $<0.3$ &   $<0.008$ &  $<1.2$    &  0.1      \\
				 PG~1426+015 & $0.43\pm0.06$ & $0.0072\pm0.0001$ &    $1.9\pm0.6$ & $0.38\pm0.02$ &  $<0.03$  &    $<1.7$ &  0.2    \\
				 PG~1411+442 & $<0.2$ & $<0.01$ & $<0.9$  & $<0.3$ &  $<0.01$  &    $<1.5$   &   0.03      \\
				 PG~0050+124 & $<0.6$ & $<0.006$ &  $<1$  & $1.47\pm0.05$ &  $0.0132\pm0.0004$   &   $3\pm1$  &  0.1 \\
				 PG~0804+761 &   $<0.2$   & $<0.01$ &  $<1.2$ & $<0.2$ &  $<0.005$  &     $<1.1$    &  0.6  \\
				 PG~1448+273 &   ... &... &  ...  & $0.480\pm0.008$ &  $0.0400\pm0.0007$  &    $1.1\pm0.4$  &  0.03  \\
				 PG~1534+580 & ... &... &  ... & $0.35\pm0.01$ &  $0.0166\pm0.0005$  &    $0.14\pm0.05$  &  0.02     \\
				 PG~1535+547 & ...& ...&  ...  & $0.32\pm0.05$ &  $0.023\pm0.003$  &   $0.24\pm0.08$   &  0.004      \\
				 PG~2214+139 &  ...   & ...&  ...  & $<0.2$ &  $<0.01$  &    $<0.5$     &  0.10      \\
				 PG~0923+129 & ...  & ...& ...   & $1.27\pm0.01$ &  $0.0470\pm0.0005$ &  $0.6\pm0.2$  &  0.03 \\
				 PG~1351+640 &...  & ...&  ...  & $1.47\pm0.03$ &  $0.0321\pm0.0006$  &   $7.3\pm2.3$   & 0.01   \\
		          PG~0007+106 & ...& ...&   ...   & $0.44\pm0.02$ &  $0.025\pm0.001$   &   $2.1\pm0.7$   &   0.2 \\	
	
	 		       \hline
			\end{tabular}%}	
			 Note.-$^{*}$Similar value was reported by \citet{Esquej14}.	
		\end{minipage}	
	\end{table}

\section{Measuring the PAH feature at 11.3 $\mu\rm{\lowercase{m}}$}
\label{analysis}

There are a variety of methods and tools to measure the PAH features \citep[e.g.,][]{Smith07, Uchida00, Peeters02, Mullaney11, Xie18}. However, they are limited to spectra having a large spectral range and clear PAH features. \citet[][]{Hernan_Caballero_Hatziminaoglou11} presented a method that implements a linear interpolation to fit a local continuum between two narrow bands on both sides of the PAH feature and integrates the area between the local continuum and spectral feature. This method allows us to measure the PAH features reliably, especially in objects with weak PAH features observed over narrow spectral ranges \citep{Esquej14}.

\subsection{The inner and larger aperture nuclear spectra} We measure the flux and EW of the PAH at 11.3 $\mu$m ($f_{11.3\,\mu\text{m}}$, $EW_{11.3\,\mu\text{m}}$) for the CC and VISIR spectra of each QSO. In order to do this, we first generate fiducial mean values of each continuum band by bootstrapping on the measured fluxes (hatched pink regions in Figure \ref{ind_nuc}). We generate 100 continuum values for the two continuum bands, taking into account the uncertainties of the spectrum. We randomly associate shorter and longer wavelength means continuum values to generate linear continua below the PAH feature. Next, to measure the flux we integrate the continuum-subtracted spectrum between the continuum bands. Then, we divide the integrated flux by the flux of the local continuum at 11.3 $\mu$m to calculate the $EW_{11.3\mu{\text m}}$. The continuum bandwidths are in the range $\Delta_{\lambda1}=10.6-11.0$ $\mu$m and $\Delta_{\lambda2}=11.4-11.7$ $\mu$m. The uncertainties are estimated repeating this process a hundred times as shown in Figure~\ref{ind_nuc}. The values reported in Table~\ref{tab:_indv_pah} are the mean and standard deviations. 

In order to determine if the PAH is clearly detected, we measure the variation of the continuum on both sidebands of the PAH feature, and use it as the level of noise. Then, comparing the flux of the PAH at 11.3 $\mu$m with the level of noise we find that for five objects the feature is clearly detected with a signal above $3{\sigma}$. For PG~0804+761 and PG~1211+143, the PAH is also detected with a signal above $3{\sigma}$. However, the large errors in the spectra, due to the marginal weather conditions during the observations \citep[precipitable water vapor $\sim8$ mm,][]{M-MP17}, do not allow to clearly see the PAH above the continuum (see panel {\sc c} in Figure~\ref{ind_nuc}), and hence we report upper limits. For the other six objects, we also report upper limits (see Table~\ref{tab:_indv_pah}).

\begin{figure}
%\hspace{-1cm}
%\centering
\begin{tabular}{cc}
	\includegraphics[scale=0.37]{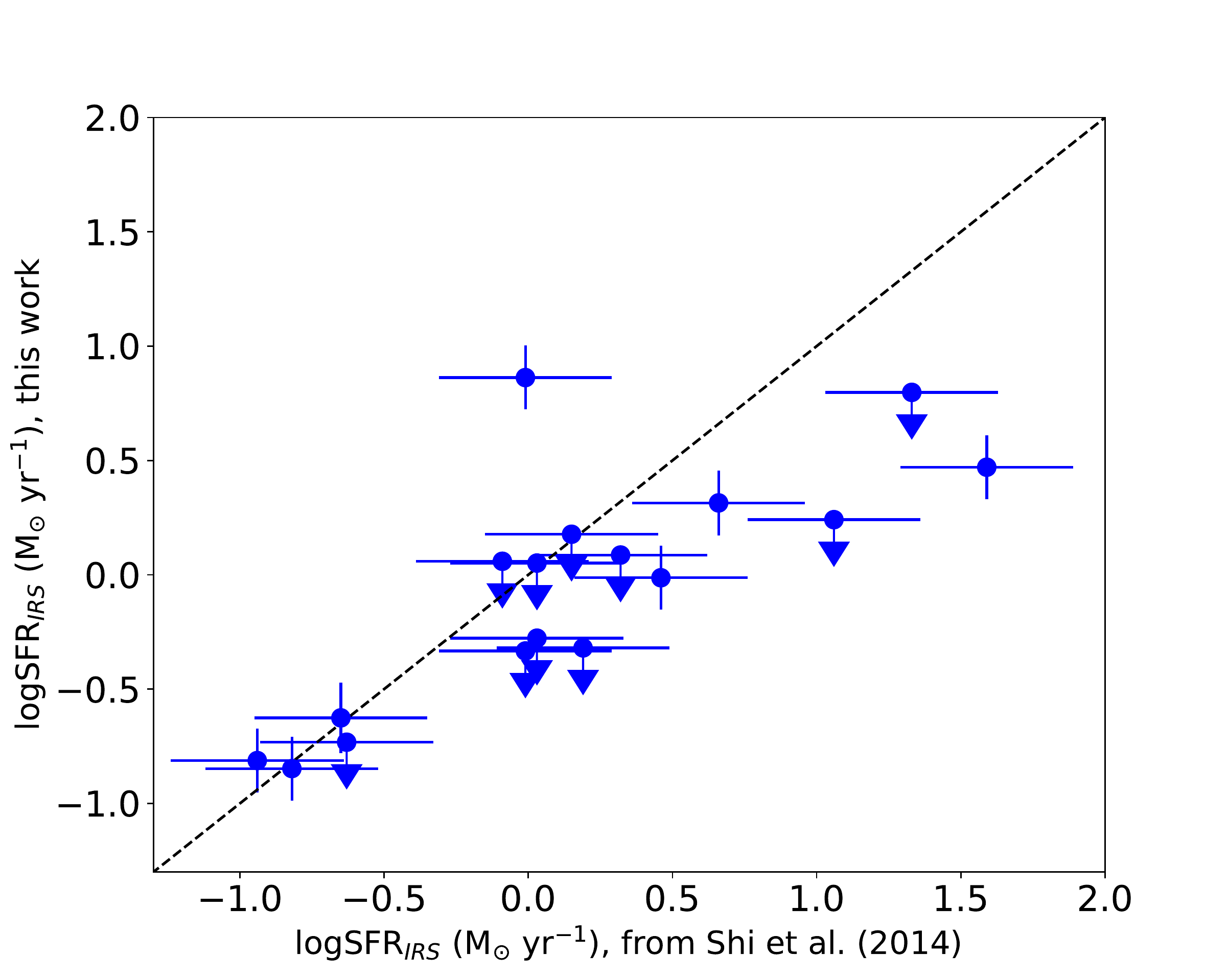} &  \includegraphics[scale=0.37]{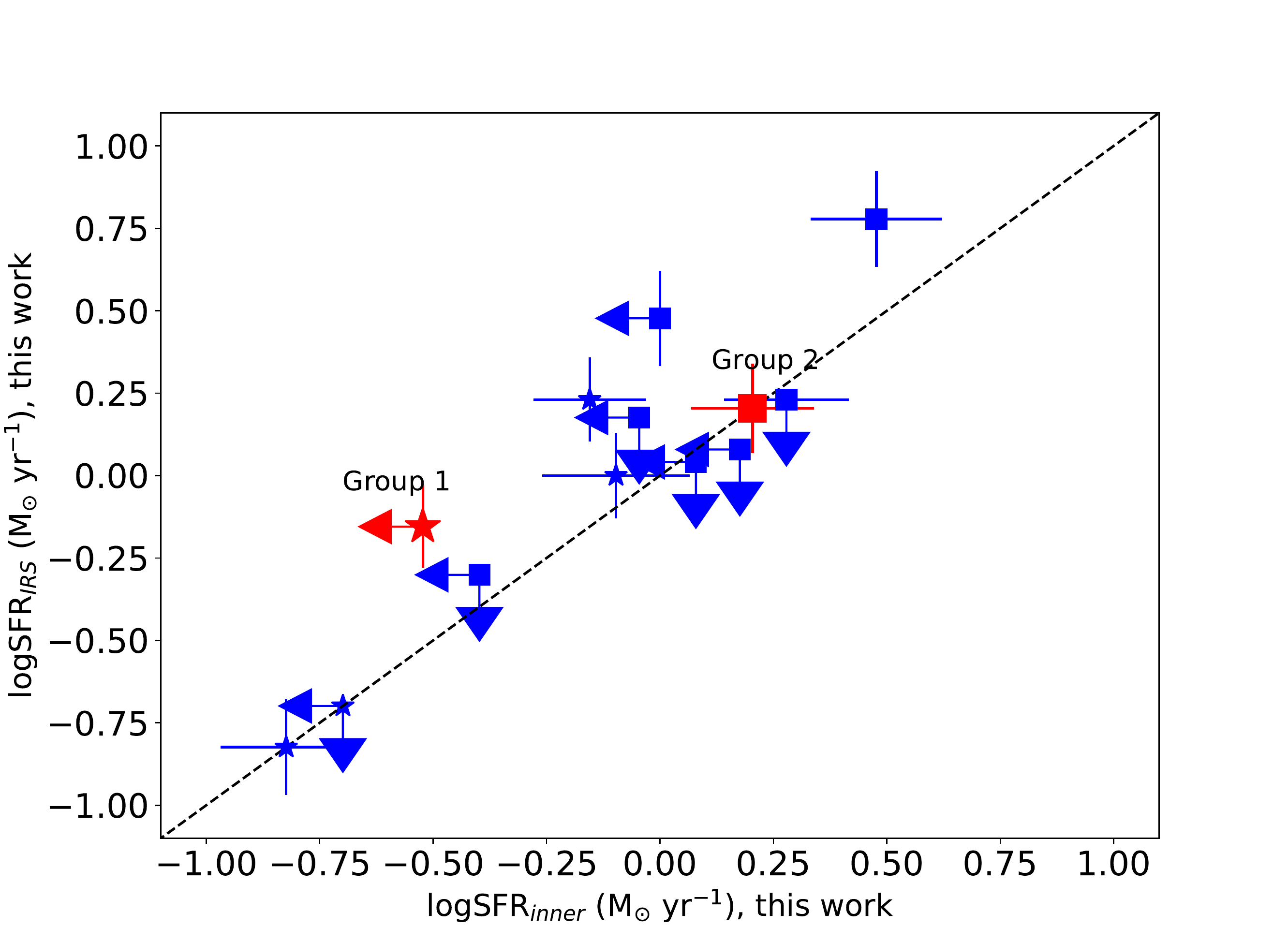} \\
\end{tabular}	
    \caption{{\bf Left:} Comparison between the larger aperture (IRS/{\it Spitzer}) SFR$_{IRS}$ estimated by \citet{Shi14} and in this work. {\bf Right:} Comparison between the larger aperture (IRS/{\it Spitzer}) SFR$_{IRS}$ and the inner SFR$_{inner}$ estimated in this work. The red star and square mark the values measured for the stacked spectra of groups 1 and 2.}
    \label{SFR_irs}
\end{figure}

In order to make a proper comparison between the results obtained from the PAH at 11.3 $\mu$m on the CC and VISIR and the IRS/{\it Spitzer} spectra, we follow the same method previously described to measure the flux and EW of the PAH at 11.3 $\mu$m for the IRS/{\it Spitzer} spectra (see Table~\ref{tab:_indv_pah}). An upper limit for the flux of the PAH is listed for those objects in which the feature is not clearly detected above the underlying continuum within the uncertainties. \citet{Shi07} uses two Drude profiles centered at 11.23 and 11.33 $\mu$m, with a fixed FWHM and the slope of the underlying silicate profile as the continuum, to measure the PAH on a sample of PG QSOs that includes most of the objects in our sample. The fluxes we measure for the PAH  feature at 11.3 $\mu$m are comparable to those reported by \citet{Shi14} (see left panel of Figure~\ref{SFR_irs}).

\begin{table}
\caption{Flux and EW of the PAH at 11.3 $\mu$m from the inner (CC and VISIR) and larger aperture (IRS/{\it Spitzer}) stacked spectra.  Column 1 lists the group. Column 2 lists the average physical scale of the inner spectra. Columns 3 and 4 give the inner (CC and VISIR) flux and EW. Column 5 gives the average physical scale of the larger Spitzer aperture, while columns 6 and 7 give the larger aperture flux and EW. \label{tab:basic2}}
	\begin{minipage}{1.\textwidth}
%\centering
\hspace{-2cm}
\scalebox{.9}{
\begin{tabular}{lcccccc}    
				\hline
				 & & \multicolumn{2}{c}{Inner aperture}&\multicolumn{2}{c}{Larger aperture}\\				\hline
				Name  & Physical scale&$f_{11.3\,\mu\text{m}}$          & EW$_{11.3\,\mu\text{m}}$ & Physical scale&$f_{11.3\,\mu\text{m}}$         & EW$_{11.3\,\mu\text{m}}$\\			                 & kpc     &$10^{-13}$  erg s$^{-1}$cm$^{-2}$& $\mu$m   & kpc   &$10^{-13}$  erg s$^{-1}$cm$^{-2}$& $\mu$m                   \\  
				\hline
				\hline				
				 Group 1 (seven QSOs)  & $<0.7$&$<0.2$ & $<0.006$ & $3.1$ & $0.56\pm0.01$ & $0.0151\pm0.0003$   \\
				 Group 2 (six QSOs) & $0.7-1$ &$0.72\pm0.03$ & $0.015\pm0.001$   & $5.2$& $0.94\pm0.02$ & $0.0214\pm0.0004$ \\	
				 All (13 QSOs)          & $<1$&$<0.09$        &  $<0.01$  & 4.1 &$0.73\pm0.01$ & $0.0179\pm0.0003$ \\		 
				\hline
			\end{tabular}}		
		\end{minipage}			
	\end{table}

\subsection{The stacked inner and larger aperture nuclear spectrum}
\label{nuc_pah}
 
The limited S/N in the CC and VISIR spectra make the detection of the PAH at 11.3 $\mu$m in individual spectra difficult. We separate the 13 QSOs with CC and VISIR spectra into two groups according to the inner physical scale ($<0.7$~kpc and $0.7$ to 1~kpc) probed by the slit-widths (see Table~\ref{tab:basic1}). This allows us to combine similar spatial scales and to keep track of the spatial resolution into the subsequent analysis, and stack the rest-frame high angular resolution spectra of each group in order to improve the S/N (see Table~\ref{tab:basic2} and Figure \ref{stack_nuc}). The stacking was done combining the normalized spectra at 10 $\mu$m of each QSO assuming an equally weighted average (see Figure \ref{stack_nuc}). We multiply the final resulting stacked spectra by their corresponding average flux at 10 $\mu$m. The final error is estimated as the error propagation of the rms uncertainty. We measure the S/N using the same continuum bands as with the individual CC and VISIR spectra. For the stacked spectrum of group 2, we obtain a S/N $=6$ that is nearly twice better than the average S/N of the group. For the stacked spectrum of group 1 we estimate a similar S/N than the average S/N of the group.

Note that group 1 includes a QSO that has an angular resolution corresponding to $376$ pc, which is nearly two times better than the average resolution of other QSOs in the group (see Table~\ref{tab:basic1}, ~\ref{tab:basic2} and Figure \ref{stack_nuc}). The result that we obtain without considering this object is the same within the uncertainties. 
We also stacked the spectra of all 13 QSOs and estimate the flux of the PAH (see Table~\ref{tab:basic2} and Figure \ref{stack_nuc}). However, this result should be used and interpreted carefully because we are combing different physical scales that range from $\sim300$ to 1 kpc. 

\begin{table}
		\caption{Inner (CC/VISIR) and larger aperture (IRS/{\it Spitzer}) star formation rates. Column 1 lists the group, column 2 and 3 give the inner and larger aperture SFRs.\label{tab:SFR_inner}}
	\begin{minipage}{1.\textwidth}
\centering
\begin{tabular}{lcc}
				\hline
				Name or group  & SFR$_{inner}$ & SFR$_{IRS}$ \\
				           &   $M_{\odot}yr^{-1}$ &$M_{\odot}yr^{-1}$ \\
				\hline
				 Group 1 (seven-QSOs)  &    $<0.3$ & $0.7\pm0.2^{*}$      \\
				 Group 2  (six-QSOs) &    $1.6\pm0.5$    & $1.6\pm0.5$   \\
				 All (13-QSOs)                &  $<0.2$ & $1.8\pm0.5$\\
				\hline        	
				\hline
			\end{tabular}			
		\end{minipage}
		Note.-$^{*}$The SFR$_{IRS}$ of group 1 is calculated with the stacked spectra of six objects, since MR~2251-178 does not have IRS/{\it Spitzer} spectrum.
	\end{table}

PG~1440+356 (MRK~478) is the only QSO exhibiting strong 11.3 $\mu$m PAH emission ($EW_{11.3\,\mu\text{m}}\sim0.04$ $\mu$m) allowing a clear detection of this feature in the CC/GTC spectrum with a S/N$=8$. We note that removing this object from group 2 gives a similar flux and the PAH is still detected. Errors in the spectra (see Figure~\ref{ind_nuc}) include only the rms uncertainty, but typical errors of nearly 10 to 15 percent associated with the photometric calibration at N-band should also be considered \citep[see,][]{A-AH16a, M-MP17}. 

 In order to make a proper comparison between the results obtained from the PAH at 11.3 $\mu$m on the CC and VISIR stacked spectra and the IRS/{\it Spitzer} spectra, we stacked the IRS/{\it Spitzer} spectra of each group and followed the same methodology described in section \ref{nuc_pah} to measure the flux and EW of the PAH at 11.3 $\mu$m (see Table~\ref{tab:basic2}). 

\section{Star formation rates}
\label{SFR}
\citet{Shipley16} calibrated the integrated luminosity of PAHs at 6.2, 7.7 and 11.3 $\mu$m as a function of the SFR using the {\it IRS/Spitzer} observations of a sample of 105 galaxies with IR luminosities between $10^{9}$ and $10^{12}$ L$_{\odot}$. They used the extinction-corrected $H_{\alpha}$ to derive the SFRs. For these calculations a Kroupa initial mass function with a slope $\alpha=2.3$ for stellar masses $0.5-100$ M$_{\odot}$ and a shallower slope $\alpha=1.3$ for the mass range $0.1-0.5$ M$_{\odot}$ was assumed \citep[][]{Kroupa03}. We calculate the SFR using their linear relationship: 

\begin{equation}
log SFR(M_{\odot} yr^{-1})=(-44.14\pm0.08)+(1.06\pm0.03)\times log L_{11.3\,\mu m} (erg s^{-1}).
\label{equ1}
\end{equation}

In Table \ref{tab:_indv_pah} we report the SFRs derived from the IRS/{\it Spitzer} spectra. \citet{Shipley16} found that a difference of a factor of two can be obtained in the SFRs derived from PAHs, according to the method used to estimate them. 

 \citet{Shi14} estimated the SFRs for most QSOs in our sample using the star forming templates from \citet{Rieke09}. However, they point out that an issue with this method is the large intrinsic scatter between PAH fluxes and SFRs, which results from the way they anchor the continuum \citep[][]{Smith07, Calzetti07}. In Figure \ref{SFR_irs} we compare the SFRs derived in this work with those derived by \citet{Shi14}. We observe that their SFRs are shifted to larger values. We find that the SFRs derived by \citet{Shi14} are on average three times larger than the SFRs we derive from the PAH at 11.3 $\mu$m using Eq.~\ref{equ1}. However, we note that if we use the PAH measurements by \citet{Shi07} and the \citet{Shipley16} relationship (Eq.~\ref{equ1}), the SFRs are similar to those derived by us (see Table~\ref{tab:_indv_pah}).

\subsection{Star formation in the central few hundred pc}

%#############################
Previous works attributed the origin of PAH features in the IRS/{\it Spitzer} spectra of local QSOs to star forming regions in the host galaxy on scales of few kiloparsecs \citep[e.g.,][]{Shi07, Schweitzer08, Shi14}. However, high angular resolution observations of local Seyferts have shown that the PAH at 11.3 $\mu$m can be detected at scales of few pc and hundreds of pc \citep{Esquej14}, supporting the idea that this feature can survive the strong nuclear radiation of the AGN \citep{Hoenig10, Diamond12, A-AH14}. 

The inner ($\sim$few hundred pc) MIR emission of QSOs is mostly dominated by an unresolved component associated with the dusty torus \citep[][]{M-MP17}. However, the emission of PAH at 11.3 $\mu$m is clearly present in the high angular resolution spectrum of five QSOs, and in the stacked spectra of group 2 on scales of $\sim1$ kpc. Indeed, the emission of the PAH at 11.3 $\mu$m is strong (S/N $=8$) in the high angular resolution spectrum of PG~1440+356 (MRK~478) on scales of $\sim0.7$ kpc. 

We estimate the inner-aperture (CC and VISIR) SFR of the 13 QSOs in our sample with high angular resolution spectra (see Table~\ref{tab:_indv_pah}), the SFR of the stacked spectra of groups 1 and 2, and of the stack of all QSOs (see Table~\ref{tab:SFR_inner}). Note that PG~1440+356 (MRK~478) is one of the objects in group 2. Removing this QSO from group 2 does not change our estimation of the SFR within the uncertainties. 

Using the same PAH and technique, we calculate the larger aperture SFRs from the IRS/{\it Spitzer} spectrum of each QSO and from the stacked spectra of groups 1 and 2. Comparing the inner (few hundred pc) and larger aperture (few kpc) SFRs we find that they are similar (see right panel in Figure~\ref{SFR_irs}). Therefore, it is likely that the star formation activity detected on larger scales is mainly concentrated within the central kpc.
Additionally, we find that at least half of the SFR estimated from the {\it IRS/Spitzer} spectrum of PG~1440+356 (MRK~478) arises from an inner region of $\sim0.7$ kpc (see Figure~\ref{ind_nuc} and \ref{tab:SFR_inner}). These results suggest that the circum-nuclear SFR in local MIR-bright QSOs is more centrally peaked than previously assumed, and they give evidence on the survival of the PAH at 11.3 $\mu$m near the strong radiation field of the AGN, on scales of few hundreds of pc.

\begin{figure*}
\centering
\includegraphics[scale=0.6]{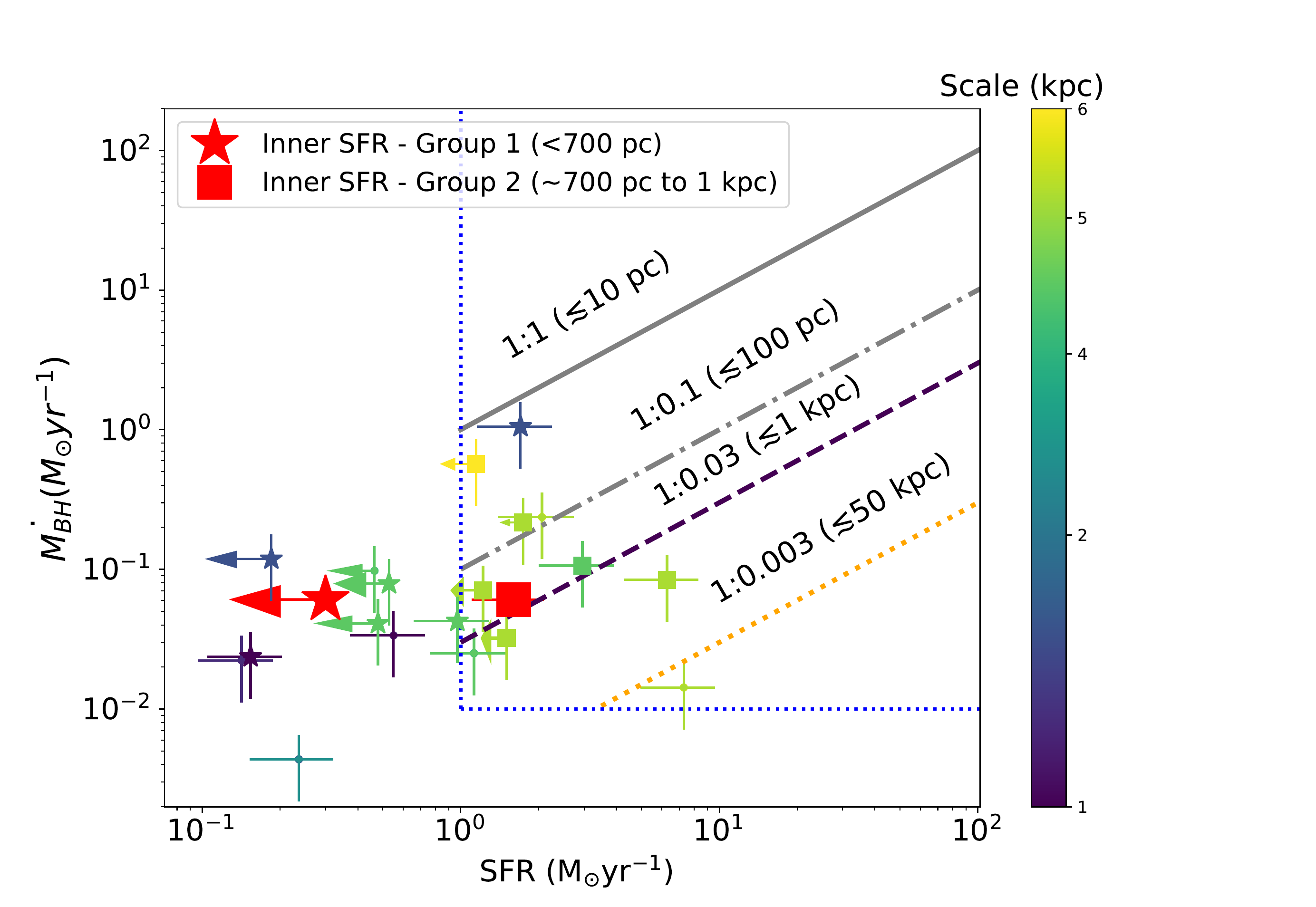} \\
              \caption{BH accretion rate as a function of the larger aperture (IRS/{\it Spitzer}) SFR of QSOs, as derived from Eq.~\ref{equ1}. The points follow the color code of the vertical bar, which indicates the central physical regions where the SFR is measured. These range from 1 to 6 kpc. The stars represent objects in group 1, while squares represent objects in group 2. The red star and square represent the inner SFR measured on the CC and VISIR stacked spectra of group 1 and 2, respectively. The uncertainty on the BH accretion rate is assumed to be 50 percent due to the large hard X-ray variability of the flux in these objects \citep[e.g.,][]{Soldi14}. The solid grey line is the 1:1 SFR:$\dot{M}_{BH}$ applicable to scales $\lesssim10$pc, the dashed-dotted grey line is the 1:0.1 ($\lesssim100$pc) ratio, the dashed dark purple line is the 1:0.03 ($\lesssim1$kpc) ratio, and the dotted orange line is the 1:0.003 relation \citep[see text and][]{Hopkins_and_ Quataert10}. The upper-right section enclosed by the dotted blue lines represents the limits above which the model is physically valid \citep[see text and][]{Hopkins_and_ Quataert10}.}
             \label{fig:acret_sfr2}
\end{figure*}

\section{The role of AGN activity on the inner star formation}
\label{discussion}
During the last decades, AGN and star formation activity have been widely studied by several authors \citep[e.g.,][]{Nicholson98, Best05, Schawinski07, Rafanelli11, Fabian12, Diniz15, Baron17}. Despite the great effort, however, the way in which AGN activity influences the gas in the host galaxy, among the various classes, sizes, and luminosities of AGN, is still uncertain \citep[][]{Volker12}.

\citet[][]{Hopkins_and_ Quataert10} presented multiscale smoothed particle hydrodynamic simulations of gravitational torques and gas inflow in AGN from kpc to sub-parsec scales, reaching a region where the material becomes a standard thin accretion disk. The simulations included as main ingredients: gas, stars, BHs, self-gravity, star formation, and stellar feedback. BH feedback was not included in order to isolate the physics of angular momentum transport. These simulations predict correlations between the BH accretion rate and the SFR at different physical scales from the central black hole. Their simulations suggest that nuclear star formation ($<10$ pc) is strongly coupled to AGN activity, following a linear relation (1:1) between BH accretion rate and SFR, while for scales $<100$ pc, $<1$ kpc and $<50$ kpc scales the relationship changes by a factor of 10, 100, and 1000, respectively. 

The AGN bolometric luminosity is estimated using the relation derived by \citet[][]{Marconi04},
\begin{equation}
\label{Lbol}
L_{bol}=k L(2-10 \text{keV}),
\end{equation}
where $k$ is a bolometric correction factor that depends on the hard X-ray luminosity ($2-10\,\text{keV}$) as a three-order polynomial, log($L$/L(2-10 \text{keV}))=1.54+0.24$\mathcal{L}$+0.012$\mathcal{L}^{2}$-0.0015$\mathcal{L}^{3}$, where $\mathcal{L}=log(L_{bol}/L_{\odot})-12$. Then, we calculate the BH accretion rate using the relation obtained by \citet[][]{Alexander_and_Hickox12},
\begin{equation}
\dot{M}_{BH} \,(M_{\odot}yr^{-1})=0.15\frac{0.1}{\epsilon}\frac{L_{bol}}{10^{45}} ,
\label{BH_ar}
\end{equation}
where $\epsilon=0.1$ is the typical value for the mass-energy conversion efficiency in the local Universe \citep[][]{Marconi04}. In Table~\ref{tab:_indv_pah} we list the BH accretion rates derived from equation~(\ref{BH_ar}). 
   
In Figure~\ref{fig:acret_sfr2} we plot the BH accretion rate against the larger aperture SFR measured in the individual IRS/{\it Spitzer} spectra of our QSOs, on scales from 1 to 6 kpc, and the inner (CC and VISIR) SFRs measured in the stacked spectra of groups 1 and 2, on scales of $<0.7$ kpc and $\sim0.7$ to 1~kpc, respectively. We compare our measurements with merger simulations of \citet[][]{Hopkins_and_ Quataert10}, and note that $\sim58$ percent of the objects in our sample lies on the valid lower BH accretion rates and SFRs of the model. We note that larger aperture SFRs measured on scales of $\sim3$ kpc for group 1 (stars) and $\sim5$ kpc for group 2 (squares) are lower than the median SFRs predicted by simulations, on the same scales and assuming a linear proportionality. A similar result was found by \citet{Ho05} using the [OII]$\lambda3727$ emission line to estimate the SFRs on similar scales for a sample of PG QSOs ($z<0.3$). \cite{Hopkins_and_ Quataert10} argued that since QSOs are on the tail of the BH accretion rate and SFR distributions, it is plausible that the AGN feedback-dominated evolutionary stage is not well accounted for in their simulations. This is especially relevant for group 1, which has an average redshift of $z=0.05$ that is 1.6 times lower than the average redshift of group 2. Therefore, we conclude that the extrapolation of the simulations is over-predicting the SFRs at the physical scales we probe for these bright nearby QSOs.

On the other hand, the low SFRs ($<0.5$ M$_{\odot}$yr$^{-1}$) measured in most QSOs of group 1 at scales of 0.7 kpc to 1 kpc (see right panel of Fig.~\ref{SFR_irs}), or at larger apertures $\sim3$ kpc (as the data in the lower left corner of Fig.~\ref{fig:acret_sfr2}), indicates that these objects are not hosted in galaxies with strong starbursts. \citet{Ho05} did not detect the on-going star formation in QSOs hosts using his [OII]$\lambda3727$ diagnosis on galactic scales either. Considering that many of these galaxies contain significant reservoirs of molecular gas \citep{Ho05, Xia12, Shangguan18}, the star formation efficiency in the host of QSO could be quenched. High spatial resolution millimeter/submillimeter interferometer could confirm this scenario. 

Another possibility is that PAH in QSOs is not effectively tracing the star formation within few kpc \citep[see e.g.,][and references therein]{Voit92, Sales10}. However, we note that for nine objects in our sample the optical SFRs (upper limits) derived by \citet{Ho05} are similar to the larger aperture PAHs-SFRs reported by us. While, the FIR-SFR derived by \citet{Petric15} for a sample of 31 local ($<0.5$) QSOs, that include the most objects in our sample, are larger. It is important to note that their FIR-SFRs are measured on even larger apertures ($\sim20$ arcsec, corresponding to several kpc) and that a fraction of the FIR emission could be heated by other sources like old stars \citep[e.g.,][]{Symeonidis16, Shangguan18}.

\begin{figure*}
\centering
\includegraphics[scale=0.6]{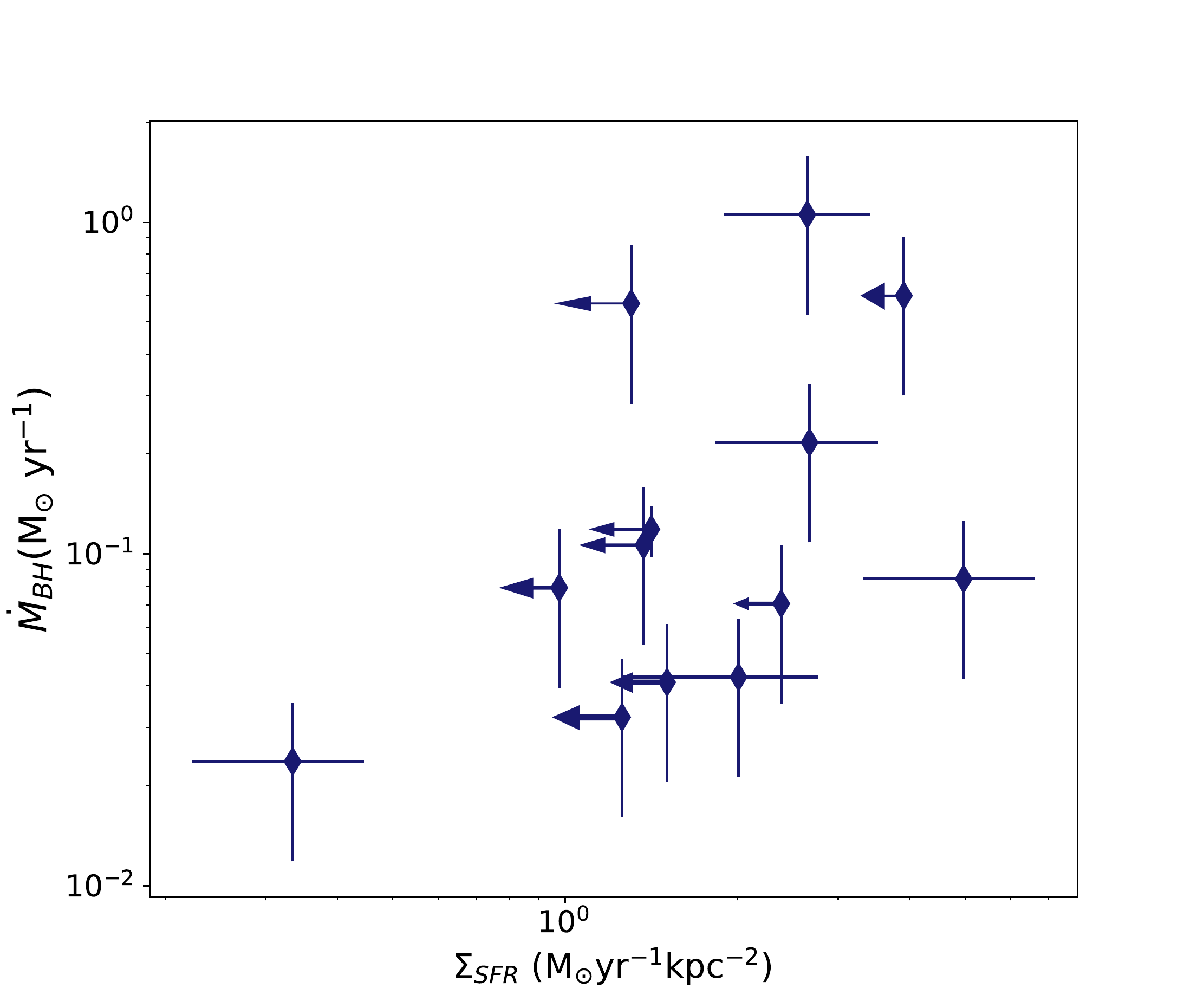} \\
              \caption{BH accretion rate as a function of the SFR density.}
             \label{fig:acret_sfrD}
\end{figure*}

For group 2 we find that the inner SFR measured in the high angular resolution stacked spectrum on scales $\lesssim1$ kpc is in agreement with the 1:0.03 SFR:$\dot{M}_{BH}$ applicable to scales of $\lesssim1$kpc. These results are consistent with a scenario in which the star formation activity in these QSOs is the same within the first 5 kpc from the central engine and is more centrally concentrated on the first kpc. 

Finally, we use our estimates of the inner SFR between $\sim0.3-1$ kpc to derive the SFR density ($\Sigma_{SFR}$), which is expected to be related to the nuclear molecular gas density of the host galaxy \citep{Schmidt59, Kennicutt89, Kennicutt98, Kennicutt_Evans12}. In Figure \ref{fig:acret_sfrD} we plot the BH accretion rate as a function of the SFR density. However, due to the large errors of the BH accretion rates and the narrow range in hard X-ray luminosity ($\sim10^{43}-10^{44}$ erg s$^{-1}$) covered by our sample, it is difficult to identify any reliable correlation. Therefore, future comparisons should include objects with lower and higher hard X-ray luminosities compared to those covered by our work. These comparisons will be useful for future spatially resolved studies of the Kennicutt-Schmidt law in quasar hosts.

More sensitive data, such as those that will be obtained with the Mid-Infrared Instrument (MIRI) on-board the {\it James Webb Space Telescope} ({\it JWST}) and higher angular resolution data to be obtained with MIR instrument on 20-40m class telescopes, will be necessary to better constrain the location of the nuclear (few pc) and circum-nuclear (few kpc) starbursts responsible for this emission, and consequently to test update models. On the other hand, measurements of the molecular gas in nearby QSOs, such as those obtained with NOEMA and ALMA, would allow the investigation of the possible correlation between inner star formation and AGN activity, since the large molecular mass detected in PG~1440+356 (MRK~478) \citep[][]{Rodriguez-Ardila03}, IZw1 \citep[][]{Evans01} and in most QSOs \citep[e.g.,][]{Maiolino97, Evans01, Scoville03, Evans06, Bertram07, Krips12, Xia12, VillarMartin13, Rodriguez14} could be protecting the surviving PAH molecules from the strong AGN radiation, as suggested by \citet{Esquej14} and \citet{A-AH14} for Seyfert galaxies.

\section{Summary and conclusions}
\label{conclusion}
 We use a sample of 20 nearby QSOs ($z<0.1$) with both {\it IRS/Spitzer} (19/20) and high angular resolution (13/20) MIR spectra in order to measure the 11.3 $\mu$m PAH emission and equivalent width on scales of a few kpc and hundreds of pc.

We detect clear 11.3 $\mu$m PAH emission in 58 percent of QSOs with IRS/{\it Spitzer} spectra, while we estimate an upper limit for the rest. We find clear emission of the PAH at 11.3 $\mu$m in five QSOs with CC and VISIR spectra, while we measure an upper limit in other eight. Additionally, we build two groups according to their physical scale probed by the slit-width ($<1$ arcsec) of the high angular resolution spectra. This allows us to combine spectra obtained on similar physical scales and to improve the S/N. For group 1 we measure the PAH at 11.3 $\mu$m on the stacked spectrum within an inner region of $<0.7$~kpc, while for the group 2 we measure the PAH on the stacked spectrum within an inner region of $\sim0.7$--1~kpc. In order to make a consistent comparison, we also stacked the IRS/{\it Spitzer} spectra of group 1 and 2, and measured the PAHs on scales of $\sim3$ and $\sim5$ kpc. 

Using the same technique and the luminosity of the PAH at 11.3 $\mu$m we estimate larger (few kpc) and inner (few hundred pc) aperture SFRs. We calculate the larger aperture SFRs for 19 QSOs with IRS/{\it Spitzer} spectra, and the inner aperture SFRs for 13 objects with high angular resolution spectra. Additionally, we calculate the larger and inner aperture SFRs for group 1 and 2 with  IRS/{\it Spitzer} and, CC and VISR stacked spectra. 

We compare the inner and larger aperture SFRs of each group and find that they are similar. We conclude that star formation activity is present in the most nearby MIR-bright QSOs at scales of a few hundred pc. Finally, we calculate both the star formation and black hole accretion rate and compare them with merger simulations. We find that QSOs in group 2 show centrally concentrated star formation activity on scales $\lesssim1$ kpc in agreement with simulations. Additionally, we note that the SFRs measured on few kpc scales are lower than those predicted by simulations. However, since QSOs are on the tail of the BH accretion rate and SFR distributions, it is possible that the evolutionary stage of objects in group 1, which have an average redshift 1.6 times lower than the average redshift of group 2 ($z=0.08$), is not well suited to these simulations. However, it is also possible that the star formation efficiency in the host galaxy is being quenched beyond the predictions of the models, and/or that the SFR derived from PAH at 11.3 $\mu$m is not well tracing the star formation activity due to the destruction or dilution of PAH by the strong radiation of the AGN.  

 In any case, future higher angular resolution and more sensitive data as those expected from the MIR instruments on 20-40m class telescopes and the Mid-Infrared Instrument (MIRI) on-board the {\it James Webb Space Telescope} ({\it JWST}) will allow us to constrain these scenarios in a better way. 

Finally, we use the inner SFR and physical scale to estimate the SFR density and compare it with the BH accretion rates. We do not find any reliable correlation. Future studies should include lower and higher hard X-ray luminosities than the ones presented here. This could help interpret future observations of the nuclear molecular gas in local QSOs.
 
\section*{Acknowledgements}

We thank the referee for an insightful report that has improved the paper significantly. This work has been partly supported by Mexican CONACyT grants CB-2011-01-167291 and CB-2016-281948. MM-P acknowledges support by UNAM-DGAPA and KASI postdoctoral fellowships. OG-M acknowledges support by the PAPIIT projects IA100516 and IA103118. AA-H. acknowledges support from the Spanish Ministry of Economy 
and Competitiveness through the grant AYA2015-64346-C2-1-P which is party funded by the FEDER program. CRA acknowledges the Ram\'on y Cajal Program of the Spanish Ministry of Economy and Competitiveness through project RYC-2014-15779 and the Spanish Plan Nacional de Astronom\'ia y Astrof\'isica under grant AYA2016-76682-C3-2-P. This work is
based on observations made with the 10.4m GTC located in the Spanish
Observatorio del Roque de Los Muchachos of the Instituto de
Astrof\'isica de Canarias, in the island La Palma. It is also based
partly on observations obtained with the \emph{Spitzer Space
  Observatory}, which is operated by JPL, Caltech, under NASA contract
1407. This research has made use of the NASA/IPAC Extragalactic
Database (NED) which is operated by JPL, Caltech, under contract with
the National Aeronautics and Space Administration. CASSIS is a
product of the Infrared Science Center at Cornell University,
supported by NASA and JPL.


\begin{thebibliography}{}
%%%%%%%%%%A
\bibitem[\protect\citeauthoryear{Alexander \& Hickox}{2012}]{Alexander_and_Hickox12} Alexander, D.~M., \& Hickox, R.~C.\ 2012, \nar, 56, 93 
\bibitem[\protect\citeauthoryear{Alonso-Herrero et al.}{2014}]{A-AH14} Alonso-Herrero, A., Ramos Almeida, C., Esquej, P., et al.\ 2014, \mnras, 443, 2766
\bibitem[\protect\citeauthoryear{Alonso-Herrero et al.}{2016a}]{A-AH16a} Alonso-Herrero, A., Esquej, P., Roche, P.~F., et al.\ 2016a, \mnras, 455, 563
\bibitem[\protect\citeauthoryear{Alonso-Herrero et al.}{2016b}]{A-AH16b} Alonso-Herrero, A., Poulton, R., Roche, P.~F., et al.\ 2016b, \mnras, 463, 2405
\bibitem[\protect\citeauthoryear{Azadi et al.}{2015}]{Azadi15} Azadi, M., Aird, J., Coil, A.~L., et al.\ 2015, \apj, 806, 187 

%%%%%%%%%%B
\bibitem[\protect\citeauthoryear{Ballantyne}{2008}]{Ballantyne08} Ballantyne, D.~R.\ 2008, \apj, 685, 787-800 
\bibitem[\protect\citeauthoryear{Baron et al.}{2017}]{Baron17} Baron, D., Netzer, H., Poznanski, D., Prochaska, J.~X., \& F{\"o}rster Schreiber, N.~M.\ 2017, \mnras, 470, 1687 
\bibitem[\protect\citeauthoryear{Beckmann \& Shrader}{2012}]{Volker12} Beckmann, V., \& Shrader, C.\ 2012, Proceedings of ``An INTEGRAL view of the high-energy sky (the first 10 years)'' - 9th INTEGRAL Workshop and celebration of the 10th anniversary of the launch (INTEGRAL 2012).~15-19 October 2012.~Bibliotheque Nationale de France, Paris, France.~Published online at \url{http://pos.sissa.it/cgi-bin/reader/conf.cgi?confid=176} \url{http://pos.sissa.it/cgi-bin/reader/conf.cgi?confid=176}, id.69, 69 
\bibitem[\protect\citeauthoryear{Bertram et al.}{2007}]{Bertram07} Bertram, T., Eckart, A., Fischer, S., et al.\ 2007, \aap, 470, 571
\bibitem[\protect\citeauthoryear{Best et al.}{2005}]{Best05} Best, P.~N., Kauffmann, G., Heckman, T.~M., et al.\ 2005, \mnras, 362, 25 
\bibitem[\protect\citeauthoryear{Burtscher et al.}{2013}]{Burtscher13} Burtscher, L., Meisenheimer, K., Tristram, K.~R.~W., et al., 2013, \aap, 558, AA149 

%%%%%%%%%%C
\bibitem[\protect\citeauthoryear{Calzetti et al.}{2007}]{Calzetti07} Calzetti, D., Kennicutt, R.~C., Engelbracht, C.~W., et al.\ 2007, \apj, 666, 870 

\bibitem[\protect\citeauthoryear{Cid Fernandes \& Terlevich}{1995}]{CidFernandez_Terlevich95} Cid Fernandes, R., Jr., \& Terlevich, R.\ 1995, \mnras, 272, 423 
\bibitem[\protect\citeauthoryear{Colina et al.}{1997}]{Colina97} Colina, L., Vargas, M.~L.~G., Delgado, R.~M.~G., et al.\ 1997, \apjl, 488, L71 

\bibitem[\protect\citeauthoryear{Cresci et al.}{2004}]{Cresci04} Cresci, G., Maiolino, R., Marconi, A., Mannucci, F., \& Granato, G.~L.\ 2004, \aap, 423, L13 
\bibitem[Cutri(1984)]{Cutri84} Cutri, R.~M.\ 1984, \baas, 16, 916 


%%%%%%%%%%D
\bibitem[\protect\citeauthoryear{Davies et al.}{2007}]{Davies07} Davies, R.~I., M{\"u}ller S{\'a}nchez, F., Genzel, R., et al.\ 2007, \apj, 671, 1388 
\bibitem[\protect\citeauthoryear{Delvecchio et al.}{2015}]{Delvecchio15} Delvecchio, I., Lutz, D., Berta, S., et al.\ 2015, \mnras, 449, 373 
\bibitem[\protect\citeauthoryear{Deo et al.}{2006}]{Deo06} Deo, R.~P., Crenshaw, D.~M., \& Kraemer, S.~B.\ 2006, \aj, 132, 321 
\bibitem[\protect\citeauthoryear{Diamond-Stanic \& Rieke}{2010}]{Diamond10} Diamond-Stanic, A.~M., \& Rieke, G.~H.\ 2010, \apj, 724, 140 
\bibitem[\protect\citeauthoryear{Diamond-Stanic \& Rieke}{2012}]{Diamond12} Diamond-Stanic, A.~M., \& Rieke, G.~H.\ 2012, \apj, 746, 168 
\bibitem[\protect\citeauthoryear{Diniz et al.}{2015}]{Diniz15} Diniz, M.~R., Riffel, R.~A., Storchi-Bergmann, T., \& Winge, C.\ 2015, \mnras, 453, 1727 
\bibitem[\protect\citeauthoryear{Dong \& Wu}{2016}]{Dong_Wu16} Dong, X.~Y., \& Wu, X.-B.\ 2016, \apj, 824, 70 
\bibitem[\protect\citeauthoryear{Du et al.}{2015}]{Du15} Du, P., Hu, C., Lu, K.-X., et al.\ 2015, \apj, 806, 22 


%%%%%%%%%%E
\bibitem[Esparza-Arredondo et al.(2018)]{Esparza17} Esparza-Arredondo, D., Gonz{\'a}lez-Mart{\'{\i}}n, O., Dultzin, D., et al.\ 2018, \apj, 859, 124 

\bibitem[\protect\citeauthoryear{Esquej et al.}{2014}]{Esquej14} Esquej, P., Alonso-Herrero, A., Gonz{\'a}lez-Mart{\'{\i}}n, O., et al.\ 2014, \apj, 780, 86 
\bibitem[\protect\citeauthoryear{Evans et al.}{2001}]{Evans01} Evans, A.~S., Frayer, D.~T., Surace, J.~A., \& Sanders, D.~B.\ 2001, \aj, 121, 1893 
\bibitem[\protect\citeauthoryear{Evans et al.}{2006}]{Evans06} Evans, A.~S., Solomon, P.~M., Tacconi, L.~J., Vavilkin, T., \& Downes, D.\ 2006, \aj, 132, 2398 

%%%%%%%%%%F

\bibitem[\protect\citeauthoryear{Fabian}{2012}]{Fabian12} Fabian, A.~C.\ 2012, \araa, 50, 455 
\bibitem[\protect\citeauthoryear{Farrah et al.}{2007}]{Farrah07} Farrah, D., Bernard-Salas, J., Spoon, H.~W.~W., et al.\ 2007, \apj, 667, 149 

\bibitem[\protect\citeauthoryear{Feruglio et al.}{2010}]{Feruglio10} Feruglio, C., Maiolino, R., Piconcelli, E., et al.\ 2010, \aap, 518, L155 


%%%%%%%%%%G
\bibitem[\protect\citeauthoryear{Gonz{\'a}lez Delgado et al.}{1998}]{Gonzalez98} Gonz{\'a}lez Delgado, R.~M., Heckman, T., Leitherer, C., et al.\ 1998, \apj, 505, 174

\bibitem[\protect\citeauthoryear{Gonz{\'a}lez-Mart{\'{\i}}n et al.}{2013}]{Gonzalez13} Gonz{\'a}lez-Mart{\'{\i}}n, O., Rodr{\'{\i}}guez-Espinosa, J.~M., D{\'{\i}}az-Santos, T., et al.\ 2013, \aap, 553, AA35 
\bibitem[\protect\citeauthoryear{Green et al.}{1986}]{Green86} Green, R.~F., Schmidt, M., \& Liebert, J.\ 1986, \apjs, 61, 305
\bibitem[\protect\citeauthoryear{G{\"u}ltekin et al.}{2009}]{Gultekin09} G{\"u}ltekin, K., Cackett, E.~M., Miller, J.~M., et al.\ 2009, \apj, 706, 404 
\bibitem[\protect\citeauthoryear{G{\"u}rkan et al.}{2015}]{Gurkan15} G{\"u}rkan, G., Hardcastle, M.~J., Jarvis, M.~J., et al.\ 2015, \mnras, 452, 3776 

%%%%%%%%%%H
\bibitem[Hao et al.(2005)]{Hao05} Hao, L., Spoon, H.~W.~W., Sloan, G.~C., et al.\ 2005, \apjl, 625, L75 

\bibitem[Heckman et al.(1997)]{Heckman97} Heckman, T.~M., Gonz{\'a}lez-Delgado, R., Leitherer, C., et al.\ 1997, \apj, 482, 114 
\bibitem[\protect\citeauthoryear{Hern{\'a}n-Caballero \& Hatziminaoglou}{2011}]{Hernan_Caballero_Hatziminaoglou11} Hern{\'a}n-Caballero, A., \& Hatziminaoglou, E.\ 2011, \mnras, 414, 500
\bibitem[\protect\citeauthoryear{H{\"o}nig et al.}{2010}]{Hoenig10} H{\"o}nig, S.~F., Kishimoto, M., Gandhi, P., et al., 2010, \aap, 515, AA23
\bibitem[Ho(2005)]{Ho05} Ho, L.~C.\ 2005, \apj, 629, 680 

\bibitem[\protect\citeauthoryear{Hopkins \& Quataert}{2010}]{Hopkins_and_ Quataert10} Hopkins, P.~F., \& Quataert, E.\ 2010, \mnras, 407, 1529 
\bibitem[\protect\citeauthoryear{Houck et al.}{2004}]{Houck04} Houck, J.~R., Roellig, T.~L., Van Cleve, J., et al.\ 2004, \procspie, 5487, 62 


%%%%%%%%%%I

\bibitem[\protect\citeauthoryear{Imanishi \& Wada}{2004}]{Imanishi_Wada04} Imanishi, M., \& Wada, K.\ 2004, \apj, 617, 214 


%%%%%%%%%%J

\bibitem[\protect\citeauthoryear{Jensen et al.}{2017}]{Jensen17} Jensen, J.~J., H{\"o}nig, S.~F., Rakshit, S., et al.\ 2017, \mnras, 470, 3071 

%%%%%%%%%%K
\bibitem[Kennicutt(1989)]{Kennicutt89} Kennicutt, R.~C., Jr.\ 1989, \apj, 344, 685 
\bibitem[Kennicutt(1998)]{Kennicutt98} Kennicutt, R.~C., Jr.\ 1998, \apj, 498, 541 
\bibitem[Kennicutt \& Evans(2012)]{Kennicutt_Evans12} Kennicutt, R.~C., \& Evans, N.~J.\ 2012, \araa, 50, 531 

\bibitem[Kim et al.(2006)]{Kim06} Kim, M., Ho, L.~C., \& Im, M.\ 2006, \apj, 642, 702 

\bibitem[\protect\citeauthoryear{Krips et al.}{2012}]{Krips12} Krips, M., Neri, R., \& Cox, P.\ 2012, \apj, 753, 135 
\bibitem[\protect\citeauthoryear{Kroupa \& Weidner}{2003}]{Kroupa03} Kroupa, P., \& Weidner, C.\ 2003, \apj, 598, 1076 

%%%%%%%%%%L

\bibitem[\protect\citeauthoryear{Lebouteiller et al.}{2011}]{Lebouteiller11} Lebouteiller, V., Barry, D.~J., Spoon, H.~W.~W., et al., 2011, \apjs, 196, 8

%%%%%%%%%%M
\bibitem[\protect\citeauthoryear{Maiolino et al.}{1997}]{Maiolino97} Maiolino, R., Thatte, N., Kroker, H., Gallimore, J.~F., \& Genzel, R.\ 1997, IAU Colloq.~159: Emission Lines in Active Galaxies: New Methods and Techniques, 113, 351 
\bibitem[\protect\citeauthoryear{Magorrian et al.}{1998}]{Magorrian98} Magorrian, J., Tremaine, S., Richstone, D., et al.\ 1998, \aj, 115, 2285 
\bibitem[\protect\citeauthoryear{Marconi et al.}{2004}]{Marconi04} Marconi, A., Risaliti, G., Gilli, R., et al.\ 2004, \mnras, 351, 169 
\bibitem[\protect\citeauthoryear{Mart{\'{\i}}nez-Paredes et al.}{2017}]{M-MP17} Mart{\'{\i}}nez-Paredes, M., Aretxaga, I., Alonso-Herrero, A., et al.\ 2017, \mnras, 468, 2 
\bibitem[\protect\citeauthoryear{Mason et al.}{2007}]{Mason07} Mason, R.~E., Levenson, N.~A., Packham, C., et al.\ 2007, \apj, 659, 241 
\bibitem[\protect\citeauthoryear{Matsuoka \& Woo}{2015}]{Matsuoka15} Matsuoka, K., \& Woo, J.-H.\ 2015, \apj, 807, 28
%\bibitem[\protect\citeauthoryear{Melnick et al.}{2015}]{Melnick15} Melnick, J., Telles, E., De Propris, R., \& Chu, Z.-H.\ 2015, \aap, 582, A37 

\bibitem[\protect\citeauthoryear{Miller et al.}{2015}]{Miller15} Miller, J.~M., Kaastra, J.~S., Miller, M.~C., et al.\ 2015, \nat, 526, 542 

\bibitem[\protect\citeauthoryear{Mullaney et al.}{2011}]{Mullaney11} Mullaney, J.~R., Alexander, D.~M., Goulding, A.~D., \& Hickox, R.~C.\ 2011, \mnras, 414, 1082 


%%%%%%%%%%N

\bibitem[\protect\citeauthoryear{Netzer et al.}{2007}]{Netzer07} Netzer, H., Lutz, D., Schweitzer, M., et al.\ 2007, \apj, 666, 806 

\bibitem[\protect\citeauthoryear{Netzer}{2009b}]{Netzer09b} Netzer, H.\ 2009, \apj, 695, 793 
\bibitem[\protect\citeauthoryear{Nicholson et al.}{1998}]{Nicholson98} Nicholson, K.~L., Reichert, G.~A., Mason, K.~O., et al.\ 1998, \mnras, 300, 893 

%%%%%%%%%%O
\bibitem[\protect\citeauthoryear{Oliva et al.}{1995}]{Oliva95} Oliva, E., Origlia, L., Kotilainen, J.~K., \& Moorwood, A.~F.~M.\ 1995, \aap, 301, 55 

%%%%%%%%%%P
\bibitem[\protect\citeauthoryear{Packham et al.}{2005a}]{Packham05a} Packham, C., Telesco, C.~M., Hough, J.~H., \& Ftaclas, C., 2005, Revista Mexicana de Astronomia y Astrofisica Conference Series, 24, 7 
\bibitem[\protect\citeauthoryear{Peeters et al.}{2002}]{Peeters02} Peeters, E., Hony, S., Van Kerckhoven, C., et al.\ 2002, \aap, 390, 1089

\bibitem[\protect\citeauthoryear{Peterson \& Wandel}{2000}]{Peterson00} Peterson, B.~M., \& Wandel, A.\ 2000, \apjl, 540, L13

\bibitem[\protect\citeauthoryear{Petric et al.}{2015}]{Petric15} Petric, A.~O., Ho, L.~C., Flagey, N.~J.~M., \& Scoville, N.~Z.\ 2015, \apjs, 219, 22 

\bibitem[\protect\citeauthoryear{Pope et al.}{2008}]{Pope08} Pope, A., Chary, R.-R., Alexander, D.~M., et al.\ 2008, \apj, 675, 1171-1193 


\bibitem[\protect\citeauthoryear{Povi{\'c} et al.}{2016}]{Povic16} Povi{\'c}, M., M{\'a}rquez, I., Netzer, H., et al.\ 2016, \mnras, 462, 2878 

%%%%%%%%%%R
\bibitem[\protect\citeauthoryear{Rafanelli et al.}{2011}]{Rafanelli11} Rafanelli, P., La Mura, G., Bindoni, D., et al.\ 2011, Baltic Astronomy, 20, 419 
\bibitem[\protect\citeauthoryear{Ramos Almeida et al.}{2014}]{Ramos14} Ramos Almeida, C., Alonso-Herrero, A., Esquej, P., et al.\ 2014, \mnras, 445, 1130 

\bibitem[\protect\citeauthoryear{Rieke et al.}{2009}]{Rieke09} Rieke, G.~H., Alonso-Herrero, A., Weiner, B.~J., et al.\ 2009, \apj, 692, 556
\bibitem[\protect\citeauthoryear{Roche et al.}{2006}]{Roche06} Roche, P.~F., Packham, C., Telesco, C.~M., et al.\ 2006, \mnras, 367, 1689 

\bibitem[\protect\citeauthoryear{Rodr{\'{\i}}guez-Ardila \& Viegas}{2003}]{Rodriguez-Ardila03} Rodr{\'{\i}}guez-Ardila, A., \& Viegas, S.~M.\ 2003, \mnras, 340, L33 
\bibitem[\protect\citeauthoryear{Rodr{\'{\i}}guez et al.}{2014}]{Rodriguez14} Rodr{\'{\i}}guez, M.~I., Villar-Mart{\'{\i}}n, M., Emonts, B., et al.\ 2014, \aap, 565, A19

\bibitem[\protect\citeauthoryear{Rosario et al.}{2012}]{Rosario12} Rosario, D.~J., Santini, P., Lutz, D., et al.\ 2012, \aap, 545, A45 
\bibitem[\protect\citeauthoryear{Rowan-Robinson}{1977}]{Rowan-Robinson77} Rowan-Robinson, M.\ 1977, \apj, 213, 635
\bibitem[\protect\citeauthoryear{Ruschel-Dutra et al.}{2017}]{Ruschel-Dutra17} Ruschel-Dutra, D., Rodr{\'{\i}}guez Espinosa, J.~M., Gonz{\'a}lez Mart{\'{\i}}n, O., Pastoriza, M., \& Riffel, R.\ 2017, \mnras, 466, 3353 


%%%%%%%%%%S
\bibitem[\protect\citeauthoryear{Sales et al.}{2010}]{Sales10} Sales, D.~A., Pastoriza, M.~G., \& Riffel, R.\ 2010, \apj, 725, 605 
\bibitem[Shangguan et al.(2018)]{Shangguan18} Shangguan, J., Ho, L.~C., \& Xie, Y.\ 2018, \apj, 854, 158 

\bibitem[\protect\citeauthoryear{Schawinski et al.}{2007}]{Schawinski07} Schawinski, K., Thomas, D., Sarzi, M., et al.\ 2007, \mnras, 382, 1415 
\bibitem[Schmidt(1959)]{Schmidt59} Schmidt, M.\ 1959, \apj, 129, 243 

\bibitem[\protect\citeauthoryear{Schweitzer et al.}{2006}]{Schweitzer06} Schweitzer, M., Lutz, D., Sturm, E., et al.\ 2006, \apj, 649, 79 

\bibitem[\protect\citeauthoryear{Schweitzer et al.}{2008}]{Schweitzer08} Schweitzer, M., Groves, B., Netzer, H., et al.\ 2008, \apj, 679, 101-117

\bibitem[\protect\citeauthoryear{Scoville et al.}{2003}]{Scoville03} Scoville, N.~Z., Frayer, D.~T., Schinnerer, E., \& Christopher, M.\ 2003, \apjl, 585, L105 
\bibitem[\protect\citeauthoryear{Shi et al.}{2007}]{Shi07} Shi, Y., Ogle, P., Rieke, G.~H., et al.\ 2007, \apj, 669, 841S
\bibitem[\protect\citeauthoryear{Shi et al.}{2014}]{Shi14} Shi, Y., Rieke, G.~H., Ogle, P.~M., Su, K.~Y.~L., \& Balog, Z.\ 2014, \apjs, 214, 23 

\bibitem[\protect\citeauthoryear{Shimizu et al.}{2015}]{Shimizu15} Shimizu, T.~T., Mushotzky, R.~F., Mel{\'e}ndez, M., Koss, M., \& Rosario, D.~J.\ 2015, \mnras, 452, 1841 
\bibitem[\protect\citeauthoryear{Shipley et al.}{2016}]{Shipley16} Shipley, H.~V., Papovich, C., Rieke, G.~H., Brown, M.~J.~I., \& Moustakas, J.\ 2016, \apj, 818, 60 
\bibitem[\protect\citeauthoryear{Siebenmorgen et al.}{2004}]{Siebenmorgen04} Siebenmorgen, R., Kr{\"u}gel, E., \& Spoon, H.~W.~W.\ 2004, \aap, 414, 123 
\bibitem[Siebenmorgen et al.(2005)]{Siebenmorgen05} Siebenmorgen, R., Haas, M., Kr{\"u}gel, E., \& Schulz, B.\ 2005, \aap, 436, L5 

\bibitem[\protect\citeauthoryear{Smith et al.}{2007}]{Smith07} Smith, J.~D.~T., Draine, B.~T., Dale, D.~A., et al.\ 2007, \apj, 656, 770 
\bibitem[\protect\citeauthoryear{Soifer et al.}{2003}]{Soifer03} Soifer, B.~T., Bock, J.~J., Marsh, K., et al.\ 2003, \aj, 126, 143 
\bibitem[\protect\citeauthoryear{Soldi et al.}{2014}]{Soldi14} Soldi, S., Beckmann, V., Baumgartner, W.~H., et al.\ 2014, \aap, 563, A57 

\bibitem[\protect\citeauthoryear{Stanley et al.}{2015}]{Stanley15} Stanley, F., Harrison, C.~M., Alexander, D.~M., et al.\ 2015, \mnras, 453, 591
\bibitem[\protect\citeauthoryear{Symeonidis et al.}{2016}]{Symeonidis16} Symeonidis, M., Giblin, B.~M., Page, M.~J., et al.\ 2016, \mnras, 459, 257 



%%%%%%%%%%T

\bibitem[\protect\citeauthoryear{Terlevich et al.}{1990}]{E.Terlevich90} Terlevich, E., D{\'{\i}}az, A.~I., \& Terlevich, R.\ 1990, \rmxaa, 21, 218 

\bibitem[\protect\citeauthoryear{Terlevich et al.}{1992}]{E.Terlevich92} Terlevich, E., Terlevich, R., \& Liaz, A.\ 1992, Relationships Between Active Galactic Nuclei and Starburst Galaxies, 31, 231 
\bibitem[\protect\citeauthoryear{Thompson et al.}{2005}]{Thompson05} Thompson, T.~A., Quataert, E., \& Murray, N.\ 2005, \apj, 630, 167 
\bibitem[\protect\citeauthoryear{Treyer et al.}{2010}]{Treyer10} Treyer, M., Schiminovich, D., Johnson, B.~D., et al.\ 2010, \apj, 719, 1191 

%%%%%%%%%%U
\bibitem[\protect\citeauthoryear{Uchida et al.}{2000}]{Uchida00} Uchida, K.~I., Sellgren, K., Werner, M.~W., \& Houdashelt, M.~L.\ 2000, \apj, 530, 817 

%%%%%%%%%%V

\bibitem[\protect\citeauthoryear{Villar-Mart{\'{\i}}n et al.}{2013}]{VillarMartin13} Villar-Mart{\'{\i}}n, M., Rodr{\'{\i}}guez, M., Drouart, G., et al.\ 2013, \mnras, 434, 978 

\bibitem[\protect\citeauthoryear{Voit}{1992b}]{Voit92} Voit, G.~M.\ 1992, Relationships Between Active Galactic Nuclei and Starburst Galaxies, 31, 87 

%%%%%%%%%%W
\bibitem[\protect\citeauthoryear{Wada \& Norman}{2002}]{Wada02} Wada, K., \& Norman, C.~A.\ 2002, \apjl, 566, L21 
\bibitem[\protect\citeauthoryear{Watabe et al.}{2008}]{Watabe08} Watabe, Y., Kawakatu, N., \& Imanishi, M.\ 2008, \apj, 677, 895-905 
\bibitem[\protect\citeauthoryear{Werner et al.}{2004}]{Werner04} Werner, M.~W., Roellig, T.~L., Low, F.~J., et al.\ 2004, \apjs, 154, 1 

%%%%%%%%%%X

\bibitem[\protect\citeauthoryear{Xia et al.}{2012}]{Xia12} Xia, X.~Y., Gao, Y., Hao, C.-N., et al.\ 2012, \apj, 750, 92 
\bibitem[Xie et al.(2018)]{Xie18} Xie, Y., Ho, L.~C., Li, A., \& Shangguan, J.\ 2018, \apj, 860, 154 


%%%%%%%%%%Z

\bibitem[\protect\citeauthoryear{Zhou \& Zhang}{2010}]{Zhou2010} Zhou, X.-L., \& Zhang, S.-N.\ 2010, \apjl, 713, L11
%External link



\end{thebibliography}
\end{document}